\documentclass[a4paper,11pt]{article}
\pdfoutput=1 % if your are submitting a pdflatex (i.e. if you have
\usepackage{jcappub} % for details on the use of the package, please
\usepackage[T1]{fontenc} % if needed

\title{\boldmath Observables for moving, stupendously charged and massive primordial black holes}

\author[a,1]{Jenny Wagner,\note{Corresponding author.}}
\affiliation[a]{Bahamas Advanced Study Institute and Conferences, \\ 4A Ocean Heights, Hill View Circle, Stella Maris, Long Island, The Bahamas}

\emailAdd{thegravitygrinch@gmail.com}

\abstract{
Stupendously large black holes exceeding $10^{11} M_\odot$ could exist, supported by recent observations of unexpectedly massive black holes at high redshifts. 
These objects may constitute a part of dark matter or even dark energy. 
One possibility to explain the cosmic accelerated expansion could be to consider charged black holes whose mutual repulsion overcomes their gravitational attraction. 
However, the extreme charge required turns these black holes into naked singularities, whose existence is questioned by the cosmic censorship hypothesis.
Since the latter is driven by theoretical assumptions, we work out the most promising observables which are least cosmology-dependent to test their existence.
We derive the electro-magnetic and gravitational lensing effects caused by such extreme objects at distances much larger than their extent to investigate possible ways for a discovery.
Restricting searches to black holes between $10^{12}$ to $10^{14} M_\odot$, we show that such objects do not cause totally disruptive catastrophes, like dissociation of neutral hydrogen clouds or proton decay induced by strong electro-magnetic fields.
Einstein rings of the order of 10'' and rotation measures of plasma clouds subject to the magnetic fields induced by the moving black holes are identified as optimum observable signatures for now. 
Future space-based black-hole telescopes will follow up on these candidates and finally check the cosmic censorship hypothesis by their strong-field strong-lensing signatures, like an additional sub-arcsecond inner Einstein ring. 
Observable effects are so surprisingly moderate that a violation of cosmic censorship is hard to detect and even explaining cosmic expansion with moving naked singularities might be possible.
}

\keywords{Dark Matter \& Dark Energy: dark energy theory, Early Universe: primordial black holes, Stars: astrophysical black holes, High Energy Astrophysics: absorption and radiation processes}

\begin{document}
\maketitle
\flushbottom

\section{Introduction}
\label{sec:introduction}

Primordial black holes (PBHs) have long been established as dark matter candidates, \cite{bib:Hawking1971, bib:Zeldovic1966}. 
Their signatures are being searched for on mass scales ranging from sub-solar masses to primordial supermassive black holes (PSMBHs) with up to $10^{11} M_\odot$, see, for instance, \cite{bib:Cappelluti2022, bib:Carr2022} or \cite{bib:Villanueva2021} for recent overviews.
As discussed in \cite{bib:Carr_stup}, so-called Stupendously LArge Black holes (SLABs) beyond $10^{11} M_\odot$ could also evolve and exist. 
Even though no clear evidence for the existence of PBHs has been found so far, they are good candidates for SLABs. 

Conventional structure growth models, which also assume black hole formation only starts at the cosmic time when the first stars were created, were recently challenged from several James Webb Space Telescope (JWST) observations of fast galaxy evolution at high redshifts, see, for instance, \cite{bib:Boylan2023, bib:Kocevski2023, bib:Labbe2023, bib:Matthee2023, bib:Pacucci2023}. 
Thus, possible black hole formation scenarios may be extended or revised alongside explanations for these early, massive galaxies.
As far as observational signatures are concerned, \cite{bib:Carr2021} and \cite{bib:Carr_stup} elaborately showed that SLABs up to $10^{16} M_\odot$ would not leave significant anisotropic imprints on the cosmic microwave background to be easily detected and that models which have limited black hole masses below $10^{11} M_\odot$ may be oversimplified. 

To extend the mass range to even higher masses, \cite{bib:Frampton2, bib:Frampton1} introduced primordial extremely massive black holes (PEMBHs)
with masses between $10^{11}$ and $10^{22} M_\odot$ and showed that they are necessary to saturate the holographic entropy bound of the visible universe.  
The latter can be considered the entropic analogue of the cosmic critical mass density. 
On the basis of a theoretical symmetry argument, one could demand that the overall cosmic entropy should be close to this entropy bound in the same way as the overall matter density is close to the critical one. 
The latter was found to apply to our universe supported by observations of the cosmic microwave background, \cite{bib:Planck2020}. 

Leaving speculations aside whether there is an upper mass limit for individual cosmic structures or not, the overall energy density of the universe is finite and therefore, the number of objects is anti-correlated with their mass. 
Yet, these objects may occur more frequently than expected, as recent observations renewed the debate whether (P)SMBHs or (P)EMBHs could not only constitute a part of dark matter but also be sources of dark energy \citep{bib:Farrah2023, bib:Frampton3, bib:Frampton4}. 
Whether or not this could be the case is still unclear. 
Criticism has been brought forward that the approach of \cite{bib:Croker2019} to couple black hole evolution to the cosmic background, which was used by \cite{bib:Farrah2023}, was self-inconsistent in its action-based ansatz, lacked an explanation to bridge the scales in size and energy density between the black holes and dark energy, and lacked a mechanism to introduce the coupling between the latter, see, for instance, \cite{bib:Mistele2023} or \cite{bib:Gaur2023}. 

Another approach was put forward by \cite{bib:Frampton3, bib:Frampton4}, taking the idea of \cite{bib:Araya2022} to sustain PBHs with charges and extending it to PEMBHs. 
To explain the accelerated cosmic expansion from within our universe, it is assumed that non-rotating, charged PEMBHs all carry charges of the same sign and overcome their mutual gravitational attraction by their Coulomb repulsion. 
This idea suffers from the fact that the black holes that satisfy the latter condition are naked singularities, as noted in \cite{bib:Frampton4}. 
Hence, their charge-to-mass ratio exceeds the limit up to which they can still have an event horizon, see also Section~\ref{sec:application_examples} for more details.
As a consequence, the creation and evolution models as outlined in \cite{bib:Araya2022} do not apply because only black holes below the extremum charge-to-mass ratio were considered.

Nevertheless, naked singularities may still have formed around the epoch of electro-weak symmetry breaking.
For instance, treating an electron classically as a charged, rotating spacetime singularity with mass, it would fall into the same category of a naked singularity due to its high charge-to-mass-ratio \citep{bib:Carter1968, bib:Burrinskii2008}. 
Accreting charges to create a naked singularity at a later time seems difficult, as charged particles in the standard model of particle physics are all massive and, for instance, \cite{bib:Wald1974} shows that massive particles can maximally create an extremum black hole with vanishing event horizon. 
Options that are left include accretion of photons that decay at almost vanishing horizon or accretion of dark-sector particles to obtain spacetimes as investigated, for instance, in \cite{bib:Morris2023}.
Concerning a stability analysis of such primordially generated naked singularities, it is also an open question whether they could evaporate. 
The Hawking temperature may be bound from below by zero for extremum black holes, as, for instance, argued in \cite{bib:Hod2018}. 
Hence, an extension of the approach is required to determine if naked singularities can evaporate in this manner. 

Despite the possibilities for their creation, the existence of such naked singularities is questioned, as summarised, for instance, in \cite{bib:Penrose1998} or \cite{bib:Singh1999}. 
Consequently, there are not many models investigating the effects on spacetime and test masses in their vicinity.
Only few works have explored observable consequences of naked singularities in general, see, for instance, \cite{bib:Sahu2012} and references therein for an overview of works on neutral naked singularities, the recent \cite{bib:Mummery2023} for a characterisation of Kerr naked singularities, and \cite{bib:Virbhadra2002} for one of the first calculations of gravitational lensing effects in the strong gravitational field for Janis-Newman-Winicour spacetimes. 
Ref.~\cite{bib:Tsukamoto2021} then developed approaches to distinguish Reissner-Nordstr\"{o}m (RN) black holes from their naked-singular counterparts based on gravitational lensing effects in the strong gravitational field.

While distinctions between standard black holes and naked singularities in terms of observable strong-field effects are not possible yet, this paper investigates observational consequences in the quasi-Newtonian regime which are entailed by the existence of such Stupendously Charged And Massive Primordial black holes, abbreviated as SCAMPs\footnote{We call them SCAMPs and not SCAMPBHs to acknowledge the fact that they are no genuine black holes anymore with a non-vanishing event horizon.}. 
In Section~\ref{sec:theoretical_derivations}, we assume that cosmic scales of interest for observations are at distances far from the these objects, such that we can approximate them as moving point charges in the late universe.
This is complementary to approaches like \cite{bib:Araya2022}, \cite{bib:Sorkin2001} or \cite{bib:Bozzola2021} which focus on the near field effects close to a black hole or coalescing black hole binaries of much smaller masses and charges. 

Our work is motivated by the fact that SCAMPs, being naked singularities of the RN metric that lack generation and evolution models, are doubted to exist.
Hence, we investigate whether they could cause disruptive observational effects or phenomena in disagreement with existing data to refute their existence on observational grounds.
To do so, in the far-field limit, we derive the acceleration that is caused by the net Coulomb repulsion of two SCAMPs exceeding their gravitational attraction. 
The electrical field and the induced magnetic field at an observer's position are calculated subsequently, employing non-relativistic Liénard-Wiechert potentials. 
We determine the effect of these electric and magnetic fields on the observed emission and absorption spectra of neutral gas clouds and the impact on observable rotation measures of ionised plasma clouds.
At last, we also determine the gravitational lensing effects caused by such SCAMPs. 
Then, Section~\ref{sec:application_examples} applies the theory developed in Section~\ref{sec:theoretical_derivations} to the example cases of SCAMPs of masses $10^{12}$ to $10^{14} M_\odot$. 
These masses are chosen because they are small enough to be found in a volume with radius of about 100~Mpc around us to allow for relatively cosmology-independent searches.
They are also the lowest masses for which the Coulomb force overcomes their mutual gravitational attraction according to the charge-to-mass relation of \cite{bib:Frampton3}, which we use as a first working hypothesis to study observable effects of SCAMPs. 
The results shown in Section~\ref{sec:application_examples} support that non-relativistic calculations are utterly sufficient for the slow motion calculated for these objects. 
To conclude, Section~\ref{sec:conclusion} summarises the surprisingly moderate impact of these extreme objects in their far field and points out that the observables derived in this paper could instead be used to find candidates for more detailed strong-field studies with upcoming telescopes like \cite{bib:Pesce2019} and \cite{bib:Novikov2021}. 

While all calculations are made for SCAMPs, they can also be applied to any non-primordial structure fulfiling the same prerequisites. 
Yet, in the cosmology constrained by current observations, it seems unlikely that objects other than SCAMPs could have grown such extremum characteristics in mass or charge and remain stable over the cosmic time available in the standard Big Bang scenario.

%%%%%%%%%%%%%%%%%%%
\section{Theoretical derivations}
\label{sec:theoretical_derivations}

We assume that the Newtonian limit of weak gravitational fields applies, i.~e.~that all distances to the SCAMPs are much larger than their extent, if one may speak of an ``extent'' in this case.
We also assume that all objects move with velocities much smaller than the speed of light $c$ in a flat space-time.
The latter assumption is corroborated by the results obtained in Section~\ref{sec:application_examples}.
If these SCAMPs are to replace dark energy in the cosmological model still under development and recently advanced by the findings of \cite{bib:Frampton4}, they are a part of the background mass density. 
Consequently, embedding them into a standard $\Lambda$-Cold-Dark-Matter ($\Lambda$CDM) cosmology to develop observables for future sky surveys is not appropriate (such an approach has been pursued by \cite{bib:Kastor1993} for extremum RN black holes). 
To avoid dependencies on any specific background cosmology, we consider distances out to redshifts which still fall under the assumption of standard flat space, or are in the linear Hubble flow at most. 
In the latter case, the linear background expansion can be set up by averaging complementary observables, as, for instance, demonstrated in \cite{bib:Wiltshire2013} for distances out to $100$~Mpc ($z \approx 0.024$).
A general framework to probe cosmology around our observation position with only a minimum amount of necessary model assumptions is being developed in \cite{bib:Heinesen2021}.
Here, we thus treat individual SCAMPs as masses on top of a flat space-time, whose distribution will be coarse-grained into a background mass density according to \cite{bib:Ellis1984} in a later work. 
\subsection{Motion caused by an excess Coulomb repulsion}
\label{sec:motion_caused}

\begin{figure}
\centering
\includegraphics[width=0.43\textwidth]{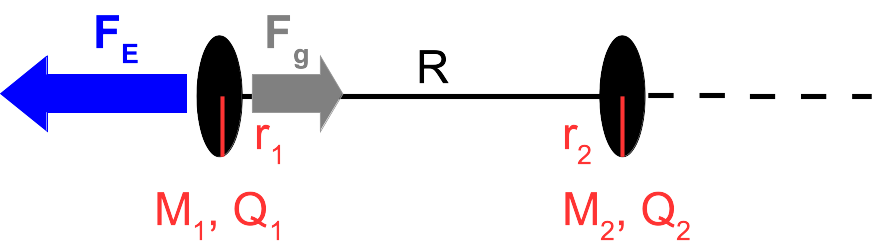}
\caption{Two charged black holes repelling each other by the Coulomb force $F_\mathrm{E}$ while being attracted towards each other by their gravitational force $F_\mathrm{g}$. Each black hole $i=1,2$ is characterised in terms of its mass $M_i$, its radial extent $r_i$, and its charge $Q_i$.}
\label{fig:scenario}
\end{figure}

Consider a scenario of two charged black holes, BH$_1$ and BH$_2$, repelling each other by a Coulomb force of strength $F_\mathrm{E}$ such that it exceeds the gravitational attraction of strength $F_\mathrm{g}$, as pictured in Fig.~\ref{fig:scenario}. 

%%%
%%%

To characterise the black holes, we use $R$ as their mutual distance and $r_i$, $M_i$, and $Q_i$ as radial extent\footnote{Usually, this is the outer event horizon of a black hole. We use the more general expression on purpose here, to be further detailed in Section~\ref{sec:application_examples} for specific cases.}, mass, and charge for black hole $i=1,2$, respectively. 
Furthermore, we assume that their extent is much smaller than their mutual distance, which amounts to the idealised approximation of point masses and charges that we use in the following to determine their dynamics. 
The repelling Coulomb force between BH$_1$ and BH$_2$ is then expressed as
\begin{equation}
F_\mathrm{E} = \dfrac{1}{4 \pi \epsilon_0} \, \dfrac{Q_1 \, Q_2}{R^2} \;,
\label{eq:FE}
\end{equation}
with $\epsilon_0$ being the vacuum permittivity. 
The gravitational force with gravitational constant $G$ is 
\begin{equation}
F_\mathrm{g} = G \, \dfrac{M_1 \, M_2}{R^2} \;, 
\label{eq:Fg}
\end{equation}
such that the ratio $\mathcal{R}$ between $F_\mathrm{g}$ and $F_\mathrm{E}$ can be defined as in \cite{bib:Frampton3}. 

\begin{figure}
\centering
\includegraphics[width=0.4\textwidth]{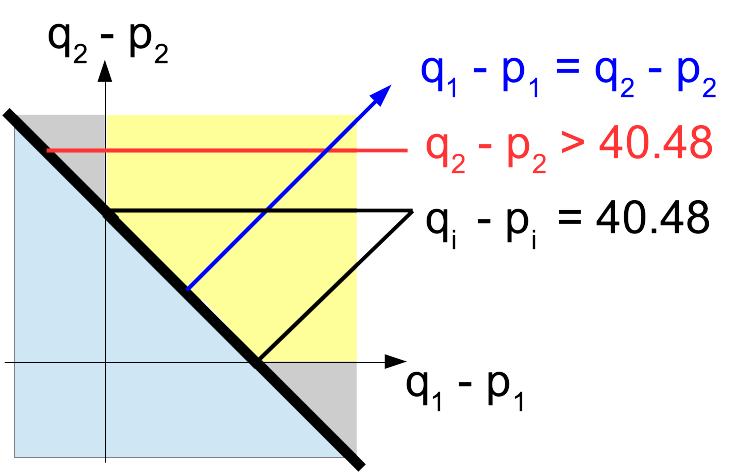}
\caption{Illustration of possible $(p_i,q_i)$-pairs, $i=1,2$, for which the Coulomb force surpasses gravitation: all pairs above the bold black line fulfil this criterion, the pairs in the blue-shaded region below do not. The blue line marks our case of two objects with identical properties. The red line visualises that one $(p_i,q_i)$ above the critical value $2 \log \left(M_\odot \sqrt{4\pi G \epsilon_0} \right) \approx 40.48$ allows for the other to have negative $(q_i-p_i)$ and enter the grey-shaded region. As detailed in Section~\ref{sec:obs_gravitational_field}, those $(p_i,q_i)$-pairs consist of a BH and a singularity, those in the yellow-shaded area consist of two singularities.}
\label{fig:bh_relations}
\end{figure}

Here, we insert the exact value for $\mathcal{R}_\mathrm{H}$ of the hydrogen atom, such that our ratio for the black holes reads
\begin{align}
\mathcal{R} &= \mathcal{R}_\mathrm{H} \left( \dfrac{M_1 M_2}{m_e m_p} \right) \left( \dfrac{Q_1 Q_2}{e^2} \right)^{-1} \\[1ex]
 &= 2.99 \times 10^{40} \times 10^{(p_1+p_2)-(q_1+q_2)} \;,
\end{align}
with $p_i$ and $q_i$ being the exponents of the masses $M_i = 10^{p_i} M_\odot$, expressed in units of solar masses $M_\odot$, and $Q_i = 10^{q_i}~\mbox{C}$, as introduced in \cite{bib:Frampton3}.
As usual, $m_e$ and $m_p$ denote the mass of the electron and the proton, respectively, and $e$ the charge of the electron.
The repulsive Coulomb force exceeds the attracting gravitation for $\mathcal{R} < 1$, which amounts to the relation
\begin{equation}
(q_1- p_1) + (q_2 - p_2) > 2 \log \left(M_\odot \sqrt{4\pi G \epsilon_0} \right) \;.
\label{eq:bh_rel}
\end{equation}
This is not only useful for our case, but also for the more general configurations of scattering charged particles off charged black holes discussed in \cite{bib:Hertzberg2023}. 
Fig.~\ref{fig:bh_relations} illustrates \eqref{eq:bh_rel} as detailed in the figure caption. 
%
%%%
%%%
%
Assuming the same properties for both BHs, \eqref{eq:bh_rel} yields for each one
\begin{equation}
q_i > p_i  + \log \left(M_\odot \sqrt{4\pi G \epsilon_0} \right) \approx p_i + 20.24 \;, \quad i = 1,2 \;.
\label{eq:qmin}
\end{equation}

Using a linear interpolation for the $Q/M$-ratio of charged black holes given in \cite{bib:Araya2022}, \cite{bib:Frampton3} derived the relation 
\begin{equation}
q_i = 2p_i + 8 + \log(4) \approx 2 p_i + 8.60
\label{eq:pg_relation}
\end{equation}
to determine the charges of SCAMPs from their given mass. 
For $p_i > \log \left(M_\odot \sqrt{4\pi G \epsilon_0} \right) - 8 + \log(4) \approx 11.63$, \eqref{eq:pg_relation} obeys $\mathcal{R} < 1$. 
As this simple interpolation may be subject to refinement in later studies, in particular due to the necessary differences in the generation, the $Q/M$-ratio does not need to obey a linear relation.
These $Q/M$-ratios should thus be considered as a working assumption to obtain first estimates as done in Section~\ref{sec:application_examples}. 

Without loss of generality, we assume that BH$_1$ is accelerated from a rest position at the origin of the coordinate system at time $t=0$ by the net force caused by the excess Coulomb repulsion of BH$_2$, i.~e.~by the force $F = (1-\mathcal{R}) F_\mathrm{E}$. 
Its acceleration is thus given by
\begin{equation}
a_1 = \dfrac{F}{M_1} = \dfrac{(1-\mathcal{R}) F_\mathrm{E}}{M_1} \;.
\label{eq:a}
\end{equation}
The motion of BH$_1$ can be observed if BH$_1$ has moved by at least $2 r_1$. 
As $r_i \ll R$, the approximation of a constant acceleration caused by the net repelling force $F$ is justified. 
The time to move by $2r_1$ from rest position is 
\begin{equation}
\delta t = \sqrt{\dfrac{2(2 r_1)}{a_1}} \;.
\label{eq:t}
\end{equation}
\subsection{Electro-magnetic fields of a moving black hole}
\label{sec:electro-magnetic_fields}

As will become clear by application examples in Section~\ref{sec:application_examples}, the black holes are moving with very small accelerations\footnote{$a_1$ will turn out to be roughly in the regime of the acceleration scale assumed for Modified Newtonian Dynamics $a_\mathrm{MOND} = 1.2\times 10^{-10}~\mbox{m}~\mbox{s}^{-2}$.}. Their final velocities are also small compared to the speed of light. 
Consequently, we can use the non-relativistic Liénard-Wiechert approach to determine the electric field and the magnetic field induced by the accelerated motion of each charged black hole.

\begin{figure}
\centering
\includegraphics[width=0.32\textwidth]{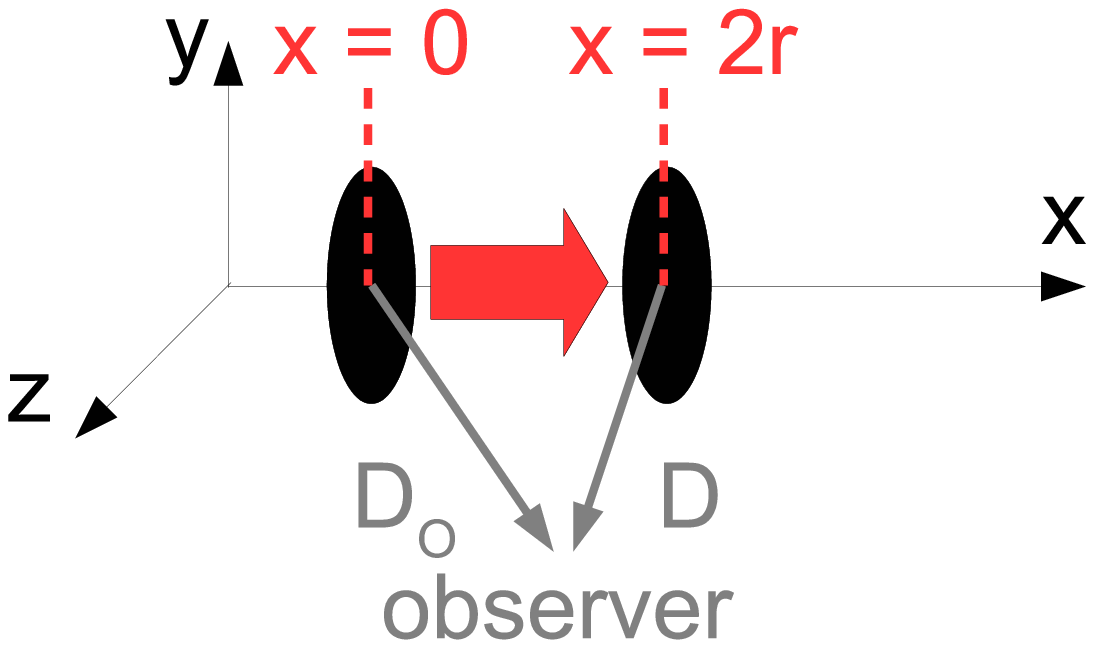} \hspace{7ex}
\includegraphics[width=0.33\textwidth]{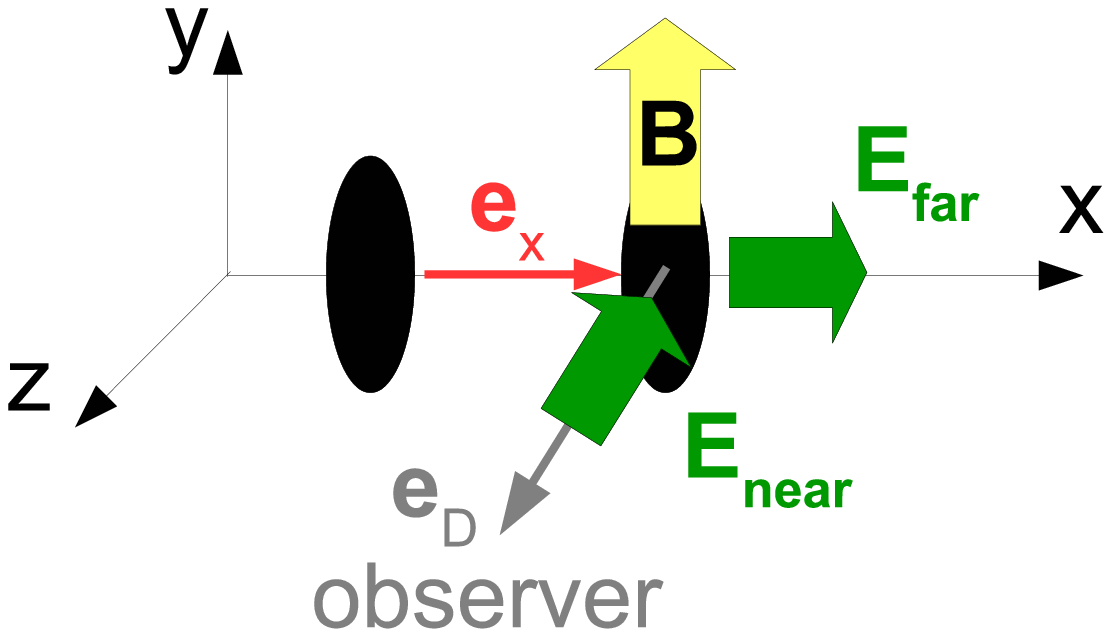}
\caption{Left: Due to the acceleration of the charged BH$_1$ caused by the net repulsion $F=F_\mathrm{E}-F_\mathrm{g}$, a magnetic field is induced in its environment. We determine the electric and the induced magnetic field at the observer position $\boldsymbol{D}_\mathrm{O} \in \mathbb{R^3}$ for a charged black hole moving along the $x$-axis from $\boldsymbol{x}=0$ at the starting time to $\boldsymbol{x}=2r_1 \boldsymbol{e}_x$ at time $\delta t$. The distance to the observer is $\boldsymbol{D}$ then and the angle enclosed between $\boldsymbol{D}$ and the $x$-axis is called $\vartheta$. Due to the finite propagation speed of electro-magnetic signals, the observer notices the fields not at $\delta t$ but at a later time $T_\mathrm{f}=\delta t + |\boldsymbol{D}|/c$. Right: Example configuration of induced $E$- and $B$-fields of a negatively charged black hole moving along the $x$-axis for an observer along the $z$-axis.}
\label{fig:B-field}
\end{figure}

Let $t$ be the time variable of the moving black hole and let the motion start at $t=0$ and $\boldsymbol{x}=0$ along the $x$-axis as shown in Fig.~\ref{fig:B-field}.
After a time $\delta t$, given by \eqref{eq:t}, has elapsed, the black hole is at position $\boldsymbol{x}=2r_1 \boldsymbol{e}_x$ on the $x$-axis, indicated by the unit vector in $x$-direction, $\boldsymbol{e}_x$.
The black hole is moving with an acceleration $\boldsymbol{a}_1$, with amplitude given by \eqref{eq:a}, at a speed of $\boldsymbol{v}_1 = \boldsymbol{a}_1 t$ in $x$-direction.
An observer at distance $\boldsymbol{D}_\mathrm{O}$ from the coordinate origin having synchronised their time coordinate $T$ at the same starting point as $t=0$ sees the motion starting at initial time $T_\mathrm{i} = \left| \boldsymbol{D}_\mathrm{O} \right| / c$ and ending at the final time $T_\mathrm{f}= \delta t + \left| \boldsymbol{D} \right|/c$ due to the finite propagation speed of electro-magnetic signals. 
A detailed derivation of the non-relativistic limit of the induced electro-magnetic fields discussed in the following subsections is shown in Appendix~\ref{app:derivation}. 

%%%
%%%
%
%
%
\subsubsection{Electric field of the moving black hole}
\label{sec:electric_field}

Based on the Li\'enard-Wiechert potentials for moving charges, the electric field as seen by the observer at distance $\boldsymbol{D}_O$ from the coordinate origin and at distance $\boldsymbol{D}$ from the moving charged black hole is given by 
\begin{equation}
\boldsymbol{E}(\boldsymbol{D}, T) = \left( \dfrac{Q_1}{4\pi \epsilon_0} \left(\dfrac{\boldsymbol{e}_D}{\left| \boldsymbol{D} \right|^2} +  \dfrac{\left( \boldsymbol{a}_1  \cdot \boldsymbol{e}_D \right) \boldsymbol{e}_{D}}{c \left| \boldsymbol{D} \right|} -\dfrac{\boldsymbol{a}_1}{c \left| \boldsymbol{D} \right|} \right) \right)_{\delta t} \;,
\label{eq:E}
\end{equation}
in which $\boldsymbol{e}_D$ denotes the unit vector in $\boldsymbol{D}$-direction and all other quantities are introduced in Section~\ref{sec:motion_caused} and evaluated at time $\delta t$ to be observed at distance $\boldsymbol{D}$ at time $T_\mathrm{f}$.
Thus, the near field is the standard Coulomb electrical field with field strength decreasing with inverse squared distance.
The far field contains two terms, both decreasing with inverse distance
\begin{eqnarray}
\boldsymbol{E}_\mathrm{near}(\boldsymbol{D}, T) &=& \left(  \dfrac{Q_1}{4\pi \epsilon_0} \dfrac{\boldsymbol{e}_D}{\left| \boldsymbol{D} \right|^2} \right)_{\delta t} \;, \label{eq:Enear} \\
\boldsymbol{E}_\mathrm{far}(\boldsymbol{D}, T) &=&  \left( \dfrac{Q_1}{4\pi \epsilon_0} \dfrac{ \left| \boldsymbol{a}_1 \right|}{c \left| \boldsymbol{D} \right|} \left( \left( \boldsymbol{e}_x  \cdot \boldsymbol{e}_D \right) \boldsymbol{e}_{D} - \boldsymbol{e}_x \right) \right)_{\delta t} \;. \label{eq:Efar}
\end{eqnarray}
The directional information in the last bracket in \eqref{eq:Efar} amounts to the unit vector orthogonal to $\boldsymbol{e}_D$ times the sine of the angle between $\boldsymbol{e}_D$ and $\boldsymbol{e}_x$, defined as $\sin(\vartheta)$ in the following.
Thus, the total amplitude of the electric field is given by $\left| \boldsymbol{E} \right| = \sqrt{ (\boldsymbol{E}_D)^2 + (\boldsymbol{E}_{D\perp})^2}$, which amounts to
\begin{equation}
\left| \boldsymbol{E}(\boldsymbol{D}, T) \right| = \left(\dfrac{\left| Q_1\right|}{4\pi \epsilon_0 \left| \boldsymbol{D} \right|^2} \sqrt{1 + \dfrac{\left| \boldsymbol{a}_1 \right|^2 \left| \boldsymbol{D} \right|^2 \sin^2(\vartheta)}{c^2}} \right)_{\delta t} \;.
\label{eq:E_abs}
\end{equation}
For distance ranges, in which the far field is not negligible compared to the near field, the maximum of the amplitude depends on $\vartheta$ and is reached for $\left| \vartheta \right| = \pi/2$, the minimum for $\vartheta = 0, \pi$. 
\subsubsection{Magnetic field induced by the moving black hole}
\label{sec:magnetic_field}

Using the same prerequisites as for the electric field, the magnetic field induced at the observer position for black hole speeds $\left| \boldsymbol{v}_1 \right| \ll c$ is 
\begin{equation}
\boldsymbol{B}(\boldsymbol{D}, T) = \left( \dfrac{\mu_0}{4\pi} Q_1 \left( \dfrac{ \boldsymbol{v}_1 \times \boldsymbol{e}_D}{\left| \boldsymbol{D} \right|^2} - \dfrac{\boldsymbol{e}_D \times \boldsymbol{a}_1}{c \left| \boldsymbol{D} \right|}\right) \right)_{\delta t} \;.
\end{equation}
Expressing velocity and acceleration in terms of amplitude and direction, we obtain
\begin{equation}
\boldsymbol{B}(\boldsymbol{D}, T) = \left( \dfrac{\mu_0}{4\pi} Q_1 \left( \dfrac{\left| \boldsymbol{v}_1 \right| }{\left| \boldsymbol{D} \right|^2} + \dfrac{\left| \boldsymbol{a}_1 \right|}{c \left| \boldsymbol{D} \right|} \right) \left( \boldsymbol{e}_x \times \boldsymbol{e}_D \right)  \right)_{\delta t}\;.
\end{equation}
With $\boldsymbol{v}_1 = \boldsymbol{a}_1 \delta t$ and \eqref{eq:t}, the simplified expression reads 
\begin{equation}
\boldsymbol{B}(\boldsymbol{D}, T) = \left( \dfrac{\mu_0}{4\pi} \dfrac{Q_1}{\left| \boldsymbol{D} \right|^2} \left( \sqrt{4r_1 \left| \boldsymbol{a}_1\right|} + \dfrac{\left| \boldsymbol{a}_1 \right| \, \left| \boldsymbol{D} \right|}{c} \right) \left( \boldsymbol{e}_x \times \boldsymbol{e}_D \right) \right)_{\delta t} \;.
\label{eq:B}
\end{equation}
Thus, the induced magnetic field splits into two parts, both pointing in the same direction orthogonal to the motion of the black hole and the direction connecting the black hole and the observer positions. The near-field part quickly drops off with the square of the distance and the far-field part only scales with inverse distance
\begin{eqnarray}
\boldsymbol{B}_\mathrm{near}(\boldsymbol{D}, T) &=& \left( \dfrac{\mu_0}{4\pi} \dfrac{Q_1 \sqrt{4r_1 \left| \boldsymbol{a}_1\right|} }{\left| \boldsymbol{D} \right|^2} \left( \boldsymbol{e}_x \times \boldsymbol{e}_D \right) \right)_{\delta t} \;, \label{eq:Bnear} \\
\boldsymbol{B}_\mathrm{far}(\boldsymbol{D}, T) &=& \left( \dfrac{\mu_0}{4\pi} \dfrac{Q_1 \left| \boldsymbol{a}_1 \right|}{c \left| \boldsymbol{D} \right|} \left( \boldsymbol{e}_x \times \boldsymbol{e}_D \right) \right)_{\delta t} \;. \label{eq:Bfar}
\end{eqnarray}
\subsection{Observable signatures}
\label{sec:observable_signatures}

The electro-magnetic fields caused by the moving charged black hole affect the observable emission and absorption spectra of gas clouds. 
As all atomic processes occur on time scales much smaller than the change of the electro-magnetic fields caused by the black hole, we assume the latter fields to be static. 
We replace the observer at distance $\boldsymbol{D}$ from the black hole by a neutral or an ionised gas cloud and investigate effects induced by the electric and the magnetic fields at this position. 
Due to the finite propagation speed of electro-magnetic signals, the fields arrive at $\boldsymbol{D}$ only at time 
\begin{equation}
T_\mathrm{f} =  \delta t + \dfrac{\left|\boldsymbol{D}\right|}{c} \;.
\label{eq:retardation}
\end{equation}
The power emitted by the moving black hole per steradian and in total is given by the Larmor formula
\begin{equation}
\dfrac{\mathrm{d} P}{\mathrm{d} \Omega} = \dfrac{Q_1^2}{4\pi c} \left| \boldsymbol{a}_ {1} - (\boldsymbol{a}_ {1} \cdot \boldsymbol{e}_D) \boldsymbol{e}_D \right|^2 \;, \quad P = \dfrac{2Q_1^2}{3c} \left| \boldsymbol{a}_ {1} \right|^2 \;.
\label{eq:larmor}
\end{equation}
One may argue that the emitted radiation could discharge the black hole.
While follow-up investigations on the stability of the charge in SCAMPs are required, we assume that the amount of radiation emitted due to the motion considered here is negligible and leaves the SCAMP invariant. 

To analyse the range of possible observable scattering properties and emission and absorption spectra of gas clouds affected by the fields of the black hole, we investigate signals traversing the affected gas clouds in different wavelength bands and relative positions between the moving black hole, the gas cloud, the signal-generating source, and the observer.  

As properties of the neutral gas cloud which is affected by the field of the black hole, we assume a HI-region consisting of atomic hydrogen at temperatures of 100~K and having a constant number density of the order of 50 atoms per cm$^3$. 
These are the properties of neutral hydrogen gas as found in the interstellar medium in the Milky Way. 
For the ionised gas cloud, specific characteristics are not detailed until Section~\ref{sec:electro-magnetic_effects} due to the large range of possibilities and many unknowns that are still subject to current research programmes.  

Assuming that SCAMPs have existed in the universe since electro-weak symmetry breaking, it is possible that charges in their environments have gathered to shield these highly charged objects and thereby reduce the fields exerted on gas clouds at farther distances.
To obtain first estimates for the magnitude of effects caused by SCAMPs we neglect shielding and treat the obtained fields induced by a SCAMP as an upper limit, which is still reducible by shielding effects. 
\subsubsection{Observable impacts of the electrical field}
\label{sec:obs_electric_field}

If the electrical fields calculated in Section~\ref{sec:electric_field} exceed 13.6~V per Bohr radius, $r_\mathrm{Bohr} = 0.0529~\mbox{nm}$, i.~e.~$2.72 \times 10^{10}~\mbox{N}~\mbox{C}^{-1}$, the electrical field of the black hole can dissociate neutral hydrogen atoms. 
Therefore, in regions with $\left| \boldsymbol{E}_\mathrm{dis} \right| \ge 2.72 \times 10^{10}~\mbox{N}~\mbox{C}^{-1}$ around the black hole, neutral hydrogen gas clouds cannot exist and no 21-cm-signals are expected to be observed from there. 

Assuming the neutral hydrogen atoms remain intact, a weaker electrical field induces a Stark effect of the bound states. 
In hydrogen, the first excited state can obtain a linear Stark effect leading to energy differences between the 2s  state, ($n=2$, $l=0$, $m=0$) in standard atomic number notation, and the highest excited 2p state ($n=2$, $l=1$, $m=0$) of $E_\mathrm{Stark} = 6 e r_\mathrm{Bohr} \left| \boldsymbol{E} \right|$.
To dominate over fine structure effects (see Section~\ref{sec:obs_magnetic_field}), the energy split caused by the electric field needs to be larger than the analogous split caused by the intrinsic magnetic fields caused by spin-orbit coupling for the fine structure and nucleus-electron coupling for the hyperfine structure. 
For the first excited state of neutral hydrogen, this implies that the energy difference between the $2p_{1/2}$ and $2p_{3/2}$ fine structure states, $4.53 \times 10^{-5}~e\mbox{V}$, should be smaller than $E_\mathrm{Stark}$. 
Hence, at least an electric field strength 
\begin{equation}
\left| \boldsymbol{E}_\mathrm{f} \right| > \dfrac{4.53 \times 10^{-5}~e\mbox{V}}{6 e r_\mathrm{Bohr}} = 1.43 \times 10^{5}~\mbox{N}~\mbox{C}^{-1}
\label{eq:Ef}
\end{equation}
is required to break the typical spectral profile of neutral hydrogen caused by the spin-orbit Zeeman effect. 
For the hyperfine structure, an analogous calculation shows that an electric field
\begin{equation}
\left| \boldsymbol{E}_\mathrm{hf} \right| > \dfrac{5.90 \times 10^{-6}~e\mbox{V}}{6 e r_\mathrm{Bohr}} = 1.86 \times 10^{4}~\mbox{N}~\mbox{C}^{-1}
\label{eq:Ehf}
\end{equation}
breaks the 21-cm line of $n=1$ in neutral hydrogen.

% http://hyperphysics.phy-astr.gsu.edu/hbase/molecule/hmol.html molecule dissociation 
% what could we say about molecular hydrogen? 

For ionised plasma clouds consisting of free electrons and protons, it could be possible that the strong electric field strength caused proton decay. 
A recent study, \cite{bib:Wistisen2021}, showed it could occur for electric field strengths about 10 times the Schwinger critical field, 
\begin{align}
\left| \boldsymbol{E}_\mathrm{proton} \right| &> 10 \left| \boldsymbol{E}_\mathrm{Schwinger} \right| \\ 
&= 10 \dfrac{2\pi m_e^2 c^3}{h e} = 1.3 \times 10^{19}~\mbox{N}~\mbox{C}^{-1} \;,
\label{eq:proton} 
\end{align}
in which $h$ is the Planck's constant. 

The polarisation induced by the electric field of a SCAMP in a plasma cloud could also be used as a signature for its presence. 
However, light emitting mechanisms of possible background sources whose light could be polarised in these plasma clouds are often unclear, such that only polarisation \emph{fluctuation} studies of observed signals traversing plasma clouds may be hints for the presence of SCAMPs on a statistical basis. 
% sources: 
% fine structure: http://hyperphysics.phy-astr.gsu.edu/hbase/quantum/hydfin.html
% hyperfine structure: https://www.feynmanlectures.caltech.edu/III_12.html
%
%
%
\subsubsection{Observable impacts of the magnetic field}
\label{sec:obs_magnetic_field}

The magnetic field strength induced at $\boldsymbol{D}$ can influence the observable spectral lines of the hydrogen atom, depending on the relative strength of this external field compared to the $\boldsymbol{B}$-field of the fine structure, or the one of the hyperfine structure.

As already noted in Section~\ref{sec:obs_electric_field}, the intrinsic Zeeman effect in the first excited state of neutral hydrogen splits the $2p_{1/2}$ and $2p_{3/2}$ states by $4.53 \times 10^{-5}~e\mbox{V}$. 
To break this typical spectral structure by an external magnetic field, the energy splitting caused by the external Zeeman effect $\Delta E_\mathrm{Z} = \mu_\mathrm{B} |\left| \boldsymbol{B} \right| \Delta m$ should exceed the intrinsic Zeeman effect. 
The difference in magnetic quantum numbers, $\Delta m$, is for the same state with quantum numbers $n, l$ and $\mu_\mathrm{B}$ is the Bohr magneton.
Thus, for magnetic fields 
\begin{equation}
\left| \boldsymbol{B}_\mathrm{f} \right| > \dfrac{4.53 \times 10^{-5}~e\mbox{V}}{2\mu_\mathrm{B}} = 0.39~\mbox{T} 
\label{eq:Bf}
\end{equation}
the spin-orbit coupling of the fine structure is broken up. 
For the hyperfine structure, an analogous calculation shows that a magnetic field
\begin{equation}
\left| \boldsymbol{B}_\mathrm{hf} \right| > \dfrac{5.90 \times 10^{-6}~e\mbox{V}}{2 \mu_\mathrm{B}} = 0.05~\mbox{T} 
\label{eq:Bhf}
\end{equation}
breaks the 21-cm-line of $n=1$ in neutral hydrogen up.

For already ionised media, like plasma clouds, we calculate the Faraday rotation of linearly polarised radiation traversing the gas cloud in the external magnetic field of a SCAMP. 
To estimate the maximum effect, we assume that all free electrons in the gas cloud are thermal electrons with negligible thermal velocity and we assume photon conservation of the radiation traversing the gas cloud and the magnetic field. 
Linearly polarised emission from quasars or fast radio bursts can serve to probe such Faraday rotation effects, see, for instance, \cite{bib:On2019} and references therein. 
If the linearly polarised radiation has a frequency much larger than the electron gyro-frequency, $\omega \gg \omega_B = eB/m_e$, and the plasma frequency, $\omega \gg \omega_\mathrm{p} = \sqrt{n_e e^2/(m_e \epsilon_0)}$, the rotation measure is given by
\begin{align}
\mathrm{RM} &= \dfrac{e^3}{8 \pi^2 \epsilon_0 m_e^2 c^3} \int \limits_{\boldsymbol{D}_\mathrm{i}}^{\boldsymbol{D}_\mathrm{f}} n_e(\boldsymbol{s}) \, \boldsymbol{B} \cdot \mathrm{d} \boldsymbol{s} \\
&= \left( \dfrac{8.37 \times 10^{-3}}{\mbox{m}^2} \right) \int \limits_{\boldsymbol{D}_\mathrm{i}}^{\boldsymbol{D}_\mathrm{f}} \left( \dfrac{n_e(\boldsymbol{s})}{1~\mbox{cm}^{-3}} \right)\left( \dfrac{\boldsymbol{B}}{1~\mbox{T}} \right)\cdot \left(\dfrac{\mathrm{d} \boldsymbol{s}}{1~\mbox{pc}} \right)\;.
\label{eq:rm}
\end{align}
The integral is taken over the traversed length $|\Delta \boldsymbol{D}|=|\boldsymbol{D}_\mathrm{f}- \boldsymbol{D}_\mathrm{i}|$ of the polarised radiation through the ionised gas cloud under the influence of the magnetic field $\boldsymbol{B}$ along the line of sight to us as observer. 
Observing linearly polarised radiation at wavelength $\lambda$, the direction of the polarisation angle $\mathrm{PA}$ is determined by
$\mathrm{PA}=\mathrm{RM}\,  \lambda^2$, such that observations over different wavelengths that show a change in the polarisation angle allow us to constrain the rotation measure. 

Yet, as also discussed, for instance, in \cite{bib:Akahori2016} or \cite{bib:Beck2011}, the quantities in the integrand of \eqref{eq:rm}, are still to be probed to much more detail with future sky surveys to reduce uncertainties in the measurements. 
Furthermore, inferring electron number densities, extensions of such plasmas and their magnetic fields is often highly model- or simulation-dependent because most observations only yield integrated quantities measured along the entire line of sight. 
Thus, identifying SCAMPs by observed rotation measures may be a difficult and degenerate endeavour, given the currently sparse knowledge on electro-magnetism in ionised gas clouds even in the absence of extremal black holes. 
%For the estimates made above with a constant $n_e = 1000~\mbox{cm}^{-3}$ and $d = 100~\mbox{pc}$ integrated over a constant magnetic field aligned parallel to the line of sight to obtain the maximum rotation measure effect, we can approximate
%\begin{equation}
%\mathrm{RM}_\mathrm{max} = 8.37 \times 10^{2} \left( \dfrac{|\boldsymbol{B}|}{1~\mbox{T}} \right) \dfrac{1}{\mbox{m}^2} \;.
%\label{eq:rm}
%\end{equation}
%
%
%
\subsubsection{Observable impacts due to gravitational lensing}
\label{sec:obs_gravitational_field}

For the sake of completeness and to establish multi-messenger probes for SCAMPs, we briefly determine the expected strong gravitational lensing signals for SCAMPs. 
Further details on a derivation and strong gravitational lensing in general, can, for instance, be found in \cite{bib:SEF}, or in the recent review \cite{bib:Wagner2019} of lens-model-independent strong gravitational lensing in the weak-field limit.
As \cite{bib:Virbhadra2002} show, the lensing signals of naked singularities are qualitatively very different from those of massive objects that have an event horizon and a photon sphere. 
Therefore, we follow their approach and derive the expected critical curves of a RN naked singularity in the strong-field regime. 

Moreover, microlensing of stars by PBHs with $M < 10^5 \, M_\odot$ is one possible signature for a detection of PBHs in this mass range. 
Observable light curves for these PBHs treated as point-like microlenses have characteristic time scales of the order of less than 100 years and are thus feasible to be acquired. 
As the characteristic time scale to observe these light curves scales with the square root of the lens mass, light curves for SCAMPs occur on time scales way beyond human life times, so that the lensing effect should be considered static, as done in this section (see Section~\ref{sec:application_examples} for further details that potential microlensing scenarios for SCAMPs considered here fall within the weak-gravitational-field limit, so that standard microlensing theory, as for instance detailed in \cite{bib:SEF} applies). 

The line element of the RN metric for an object with mass $M$ and charge $Q$ at $r=0$ is
\begin{equation}
\mathrm{d}s^2 = - \left(1 - \dfrac{r_\mathrm{S}}{r} + \dfrac{r_\mathrm{Q}^2}{r^2} \right) c^2 \mathrm{d} t^2 +  \left(1 - \dfrac{r_\mathrm{S}}{r} + \dfrac{r_\mathrm{Q}^2}{r^2} \right)^{-1} \mathrm{d} r^2 + r^2 \left( \sin^2 \theta \mathrm{d} \varphi^2 + \mathrm{d} \theta^2\right)
\label{eq:RN}
\end{equation}
in which $r_{S}=2GM/c^2$ is the Schwarzschild radius and $r_\mathrm{Q}=\sqrt{G/(4\pi\epsilon_0)}\, Q/c^2$ and the spherical coordinates are $(r, \theta, \varphi)$.
As usual, $G$ denotes the gravitational constant, $c$ the speed of light and $\epsilon_0$ the vacuum permittivity. 
The outer event horizon, if it exists, is given by
\begin{equation}
r_\mathrm{eh} = \dfrac12 r_\mathrm{S} \left(1 + \sqrt{1-\frac{4 r_\mathrm{Q}^2}{r_\mathrm{S}^2}} \right)
\label{eq:eh}
\end{equation}
and the corresponding radius of the outer photon sphere is given by
\begin{equation}
r_\mathrm{ps} = \dfrac34 r_\mathrm{S} \left( 1 + \sqrt{1-\dfrac{32}{9} \dfrac{r_\mathrm{Q}^2}{r_\mathrm{S}^2}} \right) \;.
\label{eq:rps}
\end{equation}
There are inner event horizon and photon sphere with a minus sign in front of the square-root terms, but they are not relevant here (see Appendix~\ref{app:lensing} for further details). 
RN black holes are thus turned into singularities for $2 r_\mathrm{Q} > r_\mathrm{S}$, when there is no event horizon anymore. 
Inserting the definitions of $r_\mathrm{Q}$ and $r_\mathrm{S}$ into the condition for a singularity, we arrive again at \eqref{eq:qmin}. 
According to the exact relation between $r_\mathrm{Q}$ and $r_\mathrm{S}$, there is a distinction of singularities in \emph{weakly} and \emph{strongly} naked ones, depending on whether they still have a photon sphere or not.
Strongly naked singularities having neither an event horizon nor a photon sphere occur for $2 r_\mathrm{Q} > 3/\sqrt{8} \, r_\mathrm{S} \approx 1.061 r_\mathrm{S}$. 

Analogously to \cite{bib:Virbhadra2002}, we calculate the radial geodesic of light from \eqref{eq:RN}, parametrised with an affine parameter $k$
\begin{equation}
\dfrac{{\rm{d}}^2r}{{\rm{d}}k^2} = \left( 1 - \dfrac{3r_\mathrm{S}}{r} + \dfrac{2r_\mathrm{Q}^2}{r^2}\right) r \left( \sin^2 (\vartheta) \left( \dfrac{{\rm{d}}\varphi}{{\rm{d}}k} \right)^2 +  \left( \dfrac{{\rm{d}}\vartheta}{{\rm{d}}k} \right)^2 \right)
\end{equation}
to see that the right-hand side is always positive for strongly naked singularities with $2 r_\mathrm{Q} > 3/\sqrt{8} \, r_\mathrm{S}$. 
Consequently, we expect strongly naked SCAMPs to scatter light for all impact parameters $r_\mathrm{b}>0$, also shown in Fig.~\ref{fig:RN_geometry}. 

To calculate the expected strong lensing signals in the strong-field regime, we first scale all radii to $x \equiv r/r_\mathrm{S}$ and use the general deflection angle for a central potential stated in, for instance, \cite{bib:Virbhadra2002}, to obtain 
\begin{equation}
\alpha(x_0) = 2 \int \limits_{x_0}^{+\infty} \dfrac{\mathrm{d} x}{x \sqrt{\left(\frac{x}{x_0}\right)^2 \left( 1 - \frac{1}{x_0} + \frac{x_\mathrm{Q}^2}{x_0^2} \right) - \left( 1 - \frac{1}{x} + \frac{x_Q^2}{x^2} \right)}} - \pi 
\label{eq:alpha}
\end{equation}
for the radius of closest approach $r_0$ to the BH, with $x_0 \equiv r_0/r_\mathrm{S}$ and $x_\mathrm{Q} \equiv r_\mathrm{Q}/r_\mathrm{S}$. 
For standard black holes and weakly naked singularities, trajectories of scattered or orbiting light are bounded from below by the radius of the photon sphere, implying $x_0 \ge x_\mathrm{ps}$.
Only for strongly naked singularities, there is no photon sphere, such that $x_0$ can be arbitrarily close to the singularity at the origin of the coordinate system without being devoured. 
Hence, there are no relativistic images of light captured at the radius of the photonsphere(s).
The scaled impact parameter $x_\mathrm{b} \equiv r_\mathrm{b}/r_\mathrm{S}$ for $r_0$ is given by
\begin{equation}
x_\mathrm{b} = x_0 \left(1 - \dfrac{1}{x_0} + \dfrac{x_\mathrm{Q}^2}{x_0^2} \right)^{-1/2} \;.
\label{eq:impact}
\end{equation}
From this equation, we see, that the distance of closest approach $r_0$ is equal to the impact parameter $r_\mathrm{b}$ if the former is much larger than $r_\mathrm{S}$ and $r_\mathrm{Q}$. 
Projecting the impact parameter on the sky and assuming that the light-deflecting object, i.~e.~the SCAMP, is located at an angular diameter distance $D_\mathrm{d}$ along the observers' line of sight, the relation between the angular distance on the sky $\vartheta$ and the centre of the SCAMP is given by $\text{sin}(\vartheta) = r_\mathrm{b}/D_\mathrm{d}$.

Having determined the deflection angle in \eqref{eq:alpha}, the lens equation in the strong-field regime reads
\begin{equation}
\text{tan}(\beta) = \text{tan}(\vartheta) - \dfrac{D_\mathrm{ds}}{D_\mathrm{s}} \left( \text{tan}(\vartheta) + \tan(\alpha-\vartheta)\right) \;,
\label{eq:strong_le}
\end{equation}
in which $\beta$ and $\vartheta$ denote the usual angular positions of the background source and an observed multiple image on the sky.
Both denoted by $\vartheta$, the position of the multiple images on the sky coincides with the angular position of the impact parameter projected on the sky. 
Moreover, $D_\mathrm{ds}$ and $D_\mathrm{s}$ are the angular diameter distances between the lensing object and the source and the angular diameter distance between the observer and the source, respectively, see, for instance, \cite{bib:Eiroa2002, bib:Bozza2002} for more details. 
For small angles $\beta, \vartheta$, and $\alpha$, the weak-field lens equation $\beta = \vartheta - D_\mathrm{ds}/D_\mathrm{s} \, \alpha$ is recovered.
BH lensing is then subject to all general lensing degeneracies as detailed in \cite{bib:Wagner4} and \cite{bib:Wagner6} like any other gravitational lens.

Subsequently, the critical curves are determined as those curves for which the magnification diverges, meaning
\begin{equation}
\mu^{-1} \equiv \mu_\mathrm{t}^{-1} \mu_\mathrm{r}^{-1} = \left( \dfrac{\sin (\beta)}{\sin (\theta)} \right) \left( \dfrac{{\rm d} \beta}{{\rm d} \theta} \right) \stackrel{!}{=} 0 \;.
\end{equation}
Tangential critical curves are thus obtained from $1/\mu_\mathrm{t}=0$, inserting $\beta =0$ into \eqref{eq:strong_le}, and radial critical curves are determined by setting $1/\mu_\mathrm{r}=0$, again using \eqref{eq:strong_le} in the required derivative. 
In the following, we denote the tangential and radial Einstein radii as $\vartheta_\mathrm{E,t1}$, $\vartheta_\mathrm{E,t2}$, and $\vartheta_\mathrm{r, E}$, respectively.
They are calculated numerically as further detailed in Appendix~\ref{app:lensing}. 
Due to the monotonicity of the deflection angle, RN BHs and weakly naked singularities only have a single Einstein radius. 
In contrast to that, the deflection angle of strongly naked singularities is increasing from $-\pi$ at $x_0=0$ to a maximum positive value and then decreasing again towards zero for $x_0 \rightarrow \infty$. 
Therefore, it is possible to obtain two tangential Einstein radii or none, the latter being caused by a large $r_\mathrm{Q}/r_\mathrm{s}$ ratio such that ${\rm{tan}}(\beta) > 0$ for all $x_0$.

%Assuming that SCAMPs have a spherically symmetric mass distribution with total mass $M$ up to $r_\mathrm{S}$, they can deflect light from background sources aligned behind their centre of symmetry along the line of sight to us as observers into a so-called Einstein ring. 
%Then, the radius of this Einstein ring as observed in angular coordinates on the sky is given by
%\begin{align}
%\theta_\mathrm{E} &= \sqrt{\dfrac{4GM}{c^2} \dfrac{D_\mathrm{ds}}{D_\mathrm{d} D_\mathrm{s}}} \equiv \sqrt{\dfrac{4GM}{c^2} \dfrac{1}{D_\mathrm{E}}} \\
%&=(9 \times10^{-5})'' \left( \dfrac{M}{M_\odot} \right)^{1/2} \left( \dfrac{D_\mathrm{E}}{1~\mbox{Mpc}}\right)^{-1/2}\;,
%\label{eq:theta_E}
%\end{align}
%in which $D_\mathrm{d}$, $D_\mathrm{ds}$, and $D_\mathrm{s}$ are the angular diameter distances between the SCAMP and us, the SCAMP and the light emitting background source, and the background source and us, respectively.
%For distances outside the linear Hubble flow, the angular diameter distances are dependent on a cosmological model. 
%If we assume that the SCAMP cosmology is $\Lambda$CDM-like, as found by numerous different observational probes, we can determine the Einstein radii by means of \eqref{eq:theta_E}, compare them to the largest observed ones and relate $r_\mathrm{E} \approx \theta_\mathrm{E} D_\mathrm{d}$ to $r_\mathrm{S}$.
%The latter can be considered as a scale to measure the maximum extent of the mass distribution of the SCAMP. 

As the SCAMPs considered in Section~\ref{sec:application_examples} are at least as heavy as large galaxies, their Einstein rings are expected to be large compared to lensing galaxies. 
Thus, observing Einstein rings caused by SCAMPs may be the most promising and clear signature for their existence, particularly if these SCAMPs occur as dark, isolated entities outside luminous structures. 
However, the probability of finding such a strong lensing configuration is deemed very low due to the extremely small cross section of a SCAMP as a gravitational lens. 

If a spherically symmetric SCAMP dominates the light deflection in such a strong gravitational lensing phenomenon, its tangential Einstein ring is expected to be very symmetric, in contrast to critical curves seen in galaxy clusters which are often disrupted by small-scale structure on galaxy scale. 
Analysing the possible multiple-image configurations, we also expect those generated by SCAMPs to deviate from standard configurations observed in galaxy clusters, as the latter often show an ellipsoidal symmetry in their light-deflecting mass density profile.
In \cite{bib:Griffiths2021}, an example multiple-image configuration is investigated for which simulations are set up to create it from a spherically symmetric gravitational lens or one with a less symmetric geometry. 
Based on the observed image parity, it can be clearly concluded that this configuration cannot be generated by a spherically symmetric mass density distribution, for instance, an individual SCAMP.

%%%%%%%%%%%%%%%%%%%
\section{Application examples}
\label{sec:application_examples}

\begin{figure*}
\centering
\includegraphics[width=0.49\textwidth]{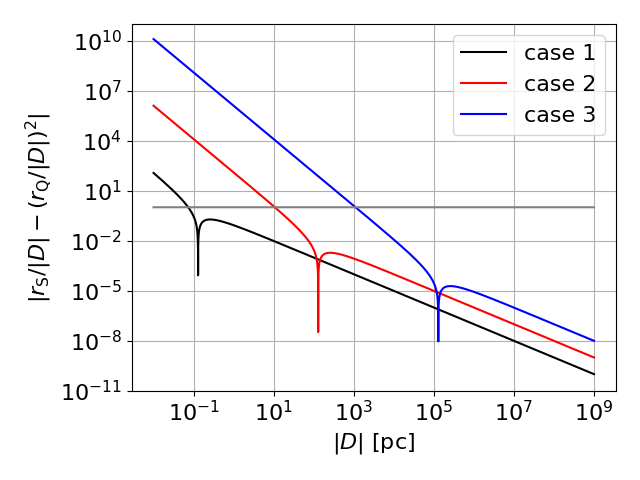}
\includegraphics[width=0.49\textwidth]{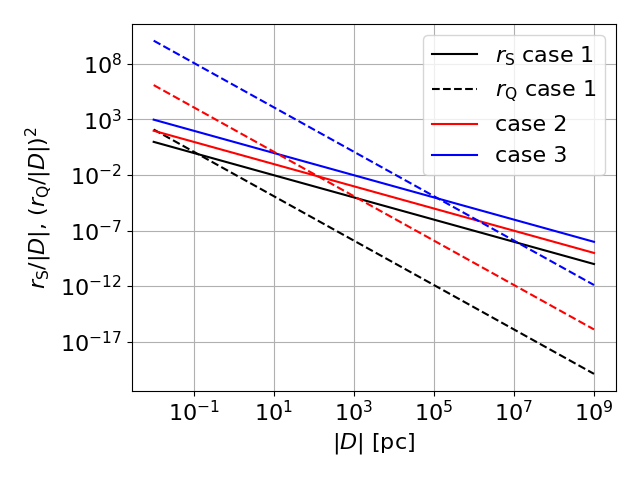}
\caption{Left: Left-hand side of \eqref{eq:approximation} versus the distance to the SCAMP. The horizontal line marks equality to one. Right: Schwarzschild (solid lines) and charge-based parts (dashed lines) of the left-hand side of \eqref{eq:approximation}. The charge-based part is always larger than the Schwarzschild one for the smallest distances, hinting at a strongly naked singularity for these kind of objects.}
\label{fig:approximation}
\end{figure*}

We use the case of two identical SCAMPs repelling each other with the specifications as denoted in Table~\ref{tab:cases} for our analysis in this section. 
First, we determine the range of validity of the Newtonian approximation by finding for which distances from a RN BH
\begin{equation}
\left| \dfrac{r_\mathrm{S}}{|\boldsymbol{D}|} - \left( \dfrac{r_\mathrm{Q}}{|\boldsymbol{D}|}\right)^2 \right|  \ll 1 \;,
\label{eq:approximation}
\end{equation}
in which $r_{S}$ and $r_\mathrm{Q}$ are given as defined after \eqref{eq:RN}.
The results are shown in Fig.~\ref{fig:approximation} (left) and the horizontal solid line marks $10^0$.  
We assume the Newtonian approximation to be valid in the regime in which the left-hand side of \eqref{eq:approximation} is smaller than 0.01.
Thus, case~1 is valid for distances $|\boldsymbol{D}| \ge 10$~pc, case~2 for $|\boldsymbol{D}| \ge100$~pc, and case~3  for $|\boldsymbol{D}| \ge 10$~kpc.
Nevertheless, in all plots of this section, distances $\left| \boldsymbol{D} \right|$ range from 0.01~pc to 1~Gpc away from the end point of the SCAMP to show all near- and far-field effects of the electro-magnetic fields. 
Retardations, $|\boldsymbol{D}|/c$ in \eqref{eq:retardation}, range from 0.03 to $3\times 10^9$ years, respectively. 

%%%
\begin{table}
\begin{center}
\begin{tabular}{cccccc}
%\hline
\textrm{Case} & $p_1 = p_2$ & $q_1=q_2$ & $r_\mathrm{S}$ & $r_\mathrm{Q}$ & $R$  \\ 
 & & & $\left[ \mbox{pc} \right]$ & $\left[ \mbox{pc} \right]$ &$\left[ \mbox{Mpc} \right]$  \\
  \noalign{\smallskip}
 \hline 
 \noalign{\smallskip}
1 & 12 & $32 + \log(4)$ & 0.10 & 0.11 & 4.0 \\
2 & 13 & $34 + \log(4)$ & 0.96 & 11.15 & 4.0 \\
3 & 14 & $36 + \log(4)$ & 9.61 & 1115.44 & 4.0\\
\noalign{\smallskip}
%\hline
\end{tabular}
\caption{\label{tab:cases} Specifications for two identical RN strongly naked singularities to be used as example SCAMPs in Section~\ref{sec:application_examples}. As defined in \protect\cite{bib:Frampton3}, $M_i = 10^{p_i} M_\odot$, $q_i= 8 + 2p_i + \log(4)$, and $Q_i = 10^{q_i}$, and $r_\mathrm{S}=2GM_i/c^2$, $r_\mathrm{Q}=\sqrt{G/(4\pi\epsilon_0)}Q_i/c^2$.}
\end{center}
\end{table}
%%%

%%%
%%% 

Fig.~\ref{fig:approximation} (right) shows the individual parts of the left-hand side of \eqref{eq:approximation}. 
For all cases, the charge-based part prevails at small distances from the black hole, while the Schwarzschild part dominates the far-field regime. 
This implies that these SCAMPs exhibit a strongly naked singularity and violate the cosmic censorship hypothesis. 
Table~\ref{tab:dynamics} summarises the resulting dynamical quantities of the moving SCAMPs. 
As can be read off Table~\ref{tab:dynamics}, the accelerations caused by the mutual repulsion are small, leading to non-relativistic motions on time scales way beyond human life times, such that we cannot observe the motions of these black holes directly. 
Yet, all motions occur on time scales much smaller than the age of the universe, such that the induced effects detailed in Section~\ref{sec:theoretical_derivations} are reasonable to consider.  
Concerning the emitted power $P$, all cases emit radiation in the range or smaller than the luminosity of an average gamma ray burst, $3 \times 10^{42}~\mbox{W}$, as determined by \cite{bib:Guetta2005}. 

%%%
\begin{table}
\begin{center}
\begin{tabular}{cccccc}
%\hline
\textrm{Case} & $\mathcal{R}$ & $\left| \boldsymbol{a}_1(\delta t)\right|$ & $\left| \boldsymbol{v}_1(\delta t)\right|$ & $\delta t$ & $P$ \\ 
 & & $\left[ \mbox{m}~\mbox{s}^{-2} \right]$ & $\left[ \mbox{m}~\mbox{s}^{-1} \right]$ & $\left[ \mbox{a} \right]$  & $\left[ \mbox{W} \right]$ \\ 
  \noalign{\smallskip}
 \hline 
 \noalign{\smallskip}
1 & $1.86 \times 10^{-1}$ & $3.8 \times 10^{-14}$ & $2.2 \times 10^{1}$  & $1.8\times10^{7}$ & $5.2 \times 10^{22}$ \\ 
2 & $1.86 \times 10^{-3}$ & $4.7 \times 10^{-11}$ & $7.6 \times 10^{2}$ & $5.1 \times 10^{5}$ & $7.8 \times 10^{32}$ \\ 
3 & $1.86 \times 10^{-5}$ & $4.7 \times 10^{-8}$ & $2.4 \times 10^{5}$  & $5.1 \times 10^{4}$ & $7.9 \times 10^{42}$ \\ 
\noalign{\smallskip}
%\hline
\end{tabular}
\caption{\label{tab:dynamics}Resulting quantities of interest for the cases summarised in Table~\ref{tab:cases}. The formulae for $\mathcal{R}$, $\boldsymbol{a}_1(\delta t)$, $\boldsymbol{v}_1(\delta t)$, and $\delta t$ are defined in Section~\ref{sec:motion_caused}, $P$ is given by \eqref{eq:larmor}.}
\end{center}
\end{table}
%%%
%
%
%
\subsection{Electro-magnetic effects}
\label{sec:electro-magnetic_effects}

\begin{figure*}
\centering
\includegraphics[width=0.49\textwidth]{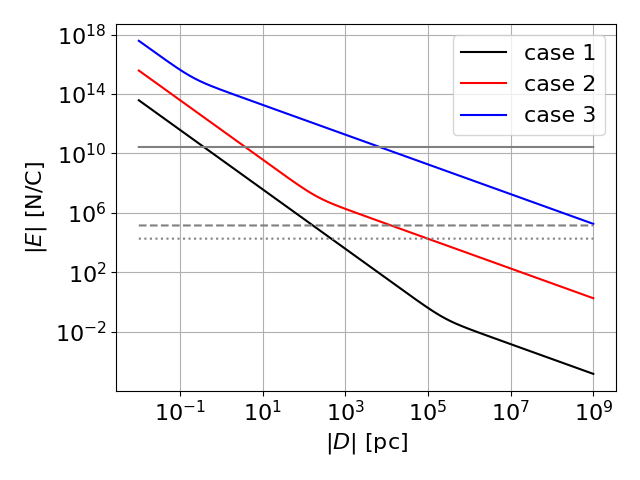}
\includegraphics[width=0.49\textwidth]{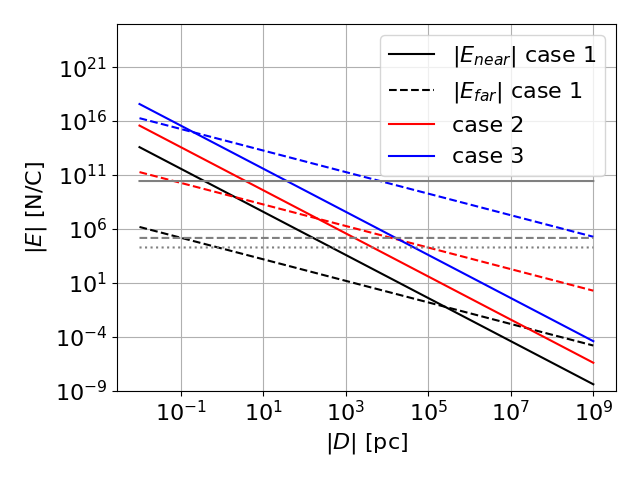}
\caption{Left: Maximal $\boldsymbol{E}$-field strength as given by \eqref{eq:E_abs} for the configurations detailed in Table~\ref{tab:cases}. Right: Near- and far-field parts as given in \eqref{eq:Enear} (solid lines) and \eqref{eq:Efar} (dashed lines) for all cases. The horizontal grey solid line marks the lowest field strength to induce dissociation of neutral hydrogen, the dashed one is the minimum $|\boldsymbol{E}|$ to break up the fine structure, \eqref{eq:Ef}, the dotted one is the minimum to break up the hyperfine structure, \eqref{eq:Ehf}.}
\label{fig:E_tot}
\end{figure*}

To probe the range of induced $\boldsymbol{E}$-fields, we plot the maximum absolute value, i.~e.~\eqref{eq:E_abs} for $\vartheta =\pi/2$ in Fig.~\ref{fig:E_tot} (left). 
Comparing the maximum values of  $|\boldsymbol{E}|$ for all distances in all cases with $|\boldsymbol{E}_\mathrm{proton}|$ of \eqref{eq:proton}, we find that the SCAMPs considered here do not induce proton decay as modelled in \cite{bib:Wistisen2021}. 
Similarly, for all distances that fulfil our approximation to the Newtonian regime, we expect neutral hydrogen not to be dissociated by the electric field of the SCAMPs considered here, either. 

Considering the fine and hyperfine structures, only SCAMPs with masses $10^{14} \, M_\odot$ break up the intrinsic emission and absorption profile of neutral hydrogen for all distances, if their motion is orthogonal to the direction to the gas cloud ($\vartheta=\pi/2$ in \eqref{eq:E_abs}). 
For cases~1 and 2, the fine and hyperfine structure can be broken up for less extremal orientations as well, but only at distances closer to the SCAMP than about 100~pc and 10~kpc, respectively. 
Thus, the impact of the electrical fields induced by these three examples of SCAMPs is expected to be perturbations for the largest part of possible distances and orientations. 

%%%
%%%

%%%
\begin{figure*}
\centering
\includegraphics[width=0.49\textwidth]{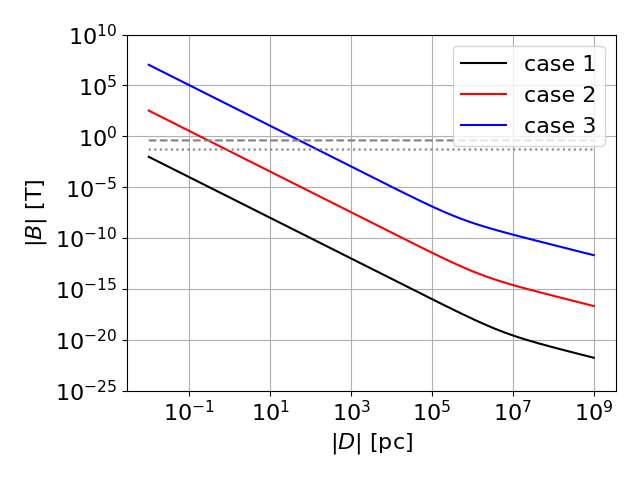}
\includegraphics[width=0.49\textwidth]{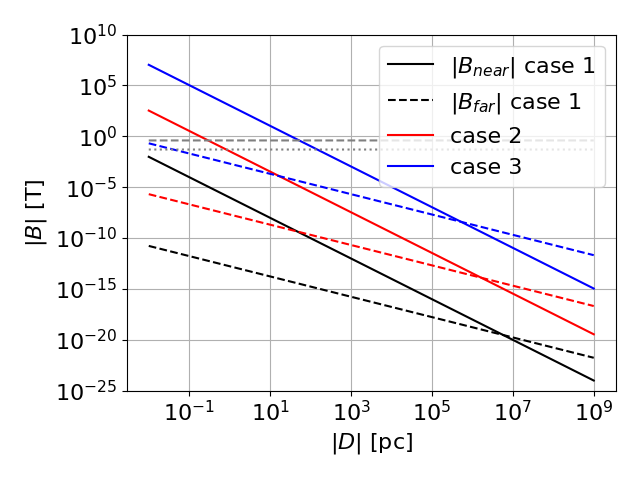}
\caption{Left: Maximal $\boldsymbol{B}$-field strength as given by \eqref{eq:B} for the configurations detailed in Table~\ref{tab:cases}. Right: Near- and far-field parts as given in \eqref{eq:Bnear} (solid lines) and \eqref{eq:Bfar} (dashed lines) for all cases. The horizontal grey dashed line is the minimum $|\boldsymbol{B}|$ to break up the fine structure, \eqref{eq:Bf}, the dotted one the minimum to break the hyperfine structure, \eqref{eq:Bhf}.}
\label{fig:B_tot}
\end{figure*}
%%%

Next, we evaluate the magnetic effects in the same manner. 
Fig.~\ref{fig:B_tot} (left) compares the maximum $\boldsymbol{B}$-field strength for all cases with each other.
Subsequently, Fig.~\ref{fig:B_tot} (right) shows the contributions of the near and the far field parts to the maximum strength for each case. 
As expected and similar to the behaviour of the electric field, for increasing mass of the SCAMP, the induced magnetic field increases and the transition from the near- to the far-field as the dominating contribution occurs at shorter distances. 
From Fig.~\ref{fig:B_tot}, we clearly see that the magnetic fields induced by the SCAMPs considered here are too weak to break the fine or hyperfine structure of neutral hydrogen for all valid distances and orientations. 

To investigate the potential contribution of SCAMPs to cosmic magnetic fields, we use the order-of-magnitude estimates as introduced in \cite{bib:Beck2011} and \cite{bib:Vallee2004}. 
Assuming the maximum magnetic field strength to be equal to these values inferred from observations, Table~\ref{tab:B-distance} summarises the distances from a SCAMP for the three cases considered here to induce such a magnetic field. 
We read off Table~\ref{tab:B-distance} that the low magnetic field strengths on cosmic scales set tight constraints on the distance to SCAMPs if we do not assume our observing position to be a fine-tuned one, such that all SCAMPs around us adopt relative orientations to suppress any induced $\boldsymbol{B}$-field. 
On galaxy and smaller scales, only the lightest SCAMPs of $10^{12} \, M_\odot$ can cause maximum $\boldsymbol{B}$-fields and still sit within the gravitationally bound structures. 
For all other black holes, the maximally induced $\boldsymbol{B}$-field exceeds the strength of other astrophysical effects causing $\boldsymbol{B}$-fields, such that specific relative orientations have to be realised to attenuate the amplitudes. 
Thus, if it is possible to overcome the practical difficulties to measure $\mathrm{RM}$s, as discussed in Section~\ref{sec:obs_magnetic_field} and also break the degeneracies between the quantities in the integrand of \eqref{eq:rm}, SCAMPs may reveal themselves in terms of their $\boldsymbol{B}$-fields induced in plasma clouds. 
Such scenarios, with black holes on smaller mass scales, are already considered for fast radio bursts with high observed rotation measures, see \cite{bib:Hilmarsson2021} for an example. 

%%%
\begin{table*}
\begin{center}
\begin{tabular}{ccccc} 
%\hline
\textrm{Scale} & $|\boldsymbol{B}|$ & $\left| \boldsymbol{D} \right|$ \textrm{case 1} & $\left| \boldsymbol{D} \right|$ \textrm{case 2} & $\left| \boldsymbol{D} \right|$ \textrm{case 3} \\ 
 & $\left[ \mbox{T} \right]$ & $\left[ \mbox{pc} \right]$ & $\left[ \mbox{kpc} \right]$ & $\left[ \mbox{Mpc} \right]$ \\ 
  \noalign{\smallskip}
 \hline 
 \noalign{\smallskip}
cosmic & $10^{-18}$ & $10^{6}$ & $> 10^{6}$ & $> 10^{3}$\\ 
\noalign{\smallskip}
galaxy & $10^{-9}$ - $10^{-8}$ & $10$ - $30$& $2$ - $6$ & $0.4$ - $2.5$ \\ 
\noalign{\smallskip}
star-burst region & $10^{-8}$ & $10$ & $2$ & $0.4$ \\ 
\noalign{\smallskip}
dense gas cloud & $10^{-8}$ - $10^{-7}$ & $< 10$ & $0.6$ - $2$ & $0.1$ - $0.4$ \\
\noalign{\smallskip}
%\hline
\end{tabular}
\caption{\label{tab:B-distance}Comparison of magnetic fields on different scales in the universe with those maximally induced at distances $|\boldsymbol{D}|$ away from the three SCAMPs detailed in Table~\ref{tab:cases}.}
\end{center}
\end{table*}
%%%
%  the potential values of rotation measures and magnetic field strengths with the estimates obtained by \cite{bib:Akahori2016} in a model-supported simulation for fast radio bursts. 
%There, it is found that the hot gas in galaxy clusters has the highest magnetic fields and rotation measures along the line of sight, $|\boldsymbol{B}| \approx 10^{-11}~\mbox{T}$ and $\mathrm{RM}\approx50~\mbox{rad}~\mbox{m}^{-2}$, while the ionised gas cloud set up as a rough estimate in Section~\ref{sec:observable_signatures} has a negligible $\mathrm{RM} < 1~\mbox{rad}~\mbox{m}^{-2}$. 
%The latter forecast is in accordance with our estimates for rotation measures induced by CPEMBHs, \eqref{eq:rm}, as well, assuming that the $|\boldsymbol{B}| \le 10^{-5}~\mbox{T}$ as induced in the quasi-Newtonian regimes of the CPEMBHs shown in Fig.~\ref{fig:B_tot}\footnote{More precisely, \cite{bib:Akahori2016} integrated along the entire line of sight, while \eqref{eq:rm} only considers the extent of the gas cloud, }.
%However, as also discussed in \cite{bib:Akahori2016}, the number density of free charges and the extension of their distribution are still very uncertain and subject of many research projects to this day. 
%We therefore consider another observation, discussed in \cite{bib:Hilmarsson2021}, which considers the proximity to a black hole as one possible cause for the large rotation measure, $\mathrm{RM}=$, and the inferred magnetic field strengths of 
%
%
%
\subsection{Gravitational lensing effects}
\label{sec:gravitational_effects}

To determine the number and sizes of the Einstein radii for the three SCAMPs of Table~\ref{tab:cases} by means of \eqref{eq:strong_le}, we additionally have to choose distances for the SCAMPs as lenses and for the background sources. 
We consider two scenarios. 
In the first, we place the SCAMP at $D_\mathrm{d} = 100$~Mpc ($z_\mathrm{d}=0.024$) as the outmost possible position in the linear Hubble flow and assume a $\Lambda$CDM-like cosmology to investigate strong lensing effects on cosmic scales for sources at typical lensing redshifts $z_\mathrm{s}=0.05$ to 1.0.
In the second scenario, we place the background source at $D_\mathrm{s} = 100$~Mpc ($z_\mathrm{s}=0.024$) and investigate the strong lensing effects for SCAMPs at distances $D_\mathrm{d}=1$ to 99~Mpc. 
%(We choose to start placing the SCAMPs at 1~Mpc to guarantee for us as observers to be in the Newtonian regime for all three CPEMBH masses.)

\begin{figure*}
\centering
\includegraphics[width=0.48\textwidth]{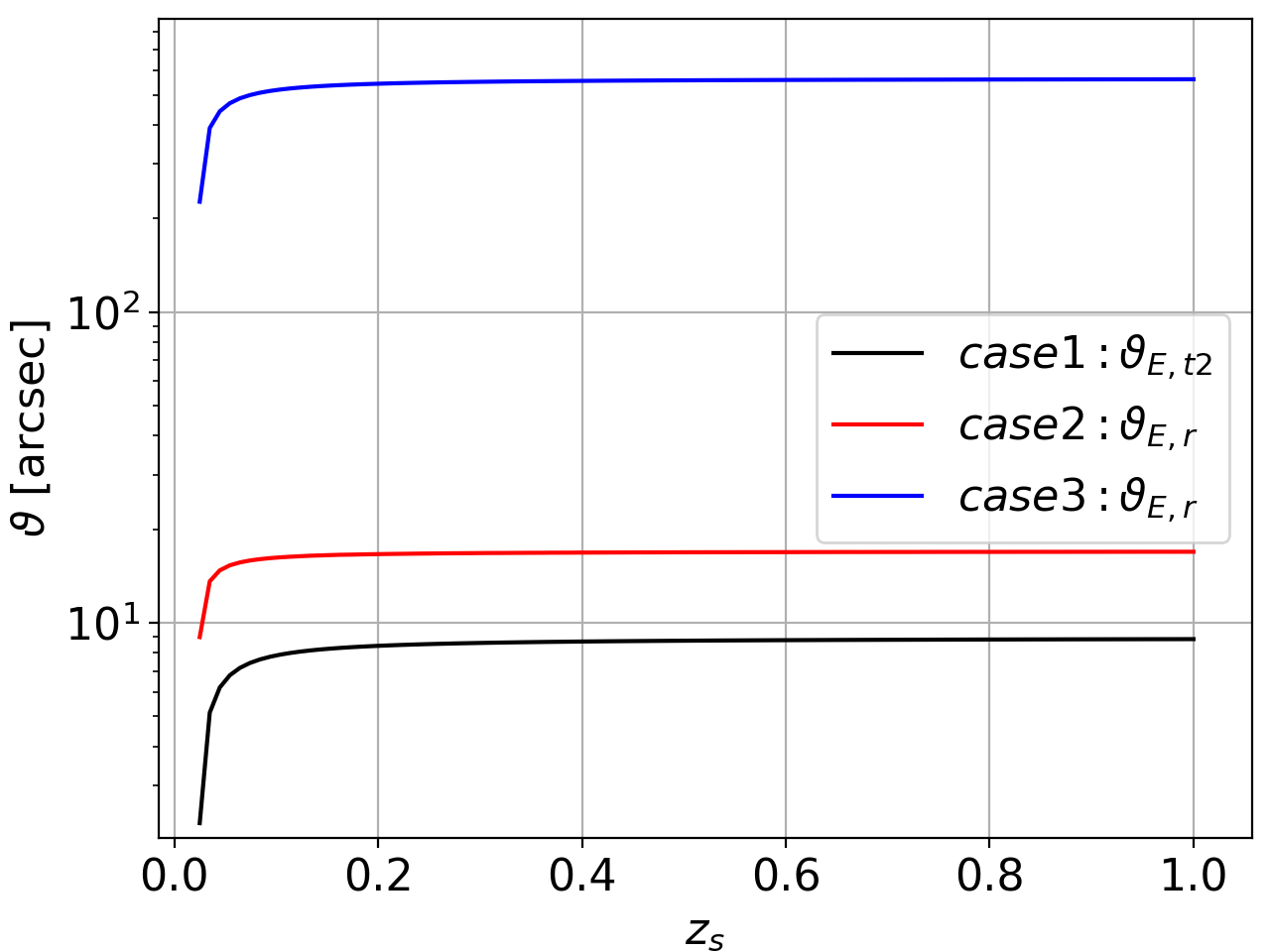}
\includegraphics[width=0.48\textwidth]{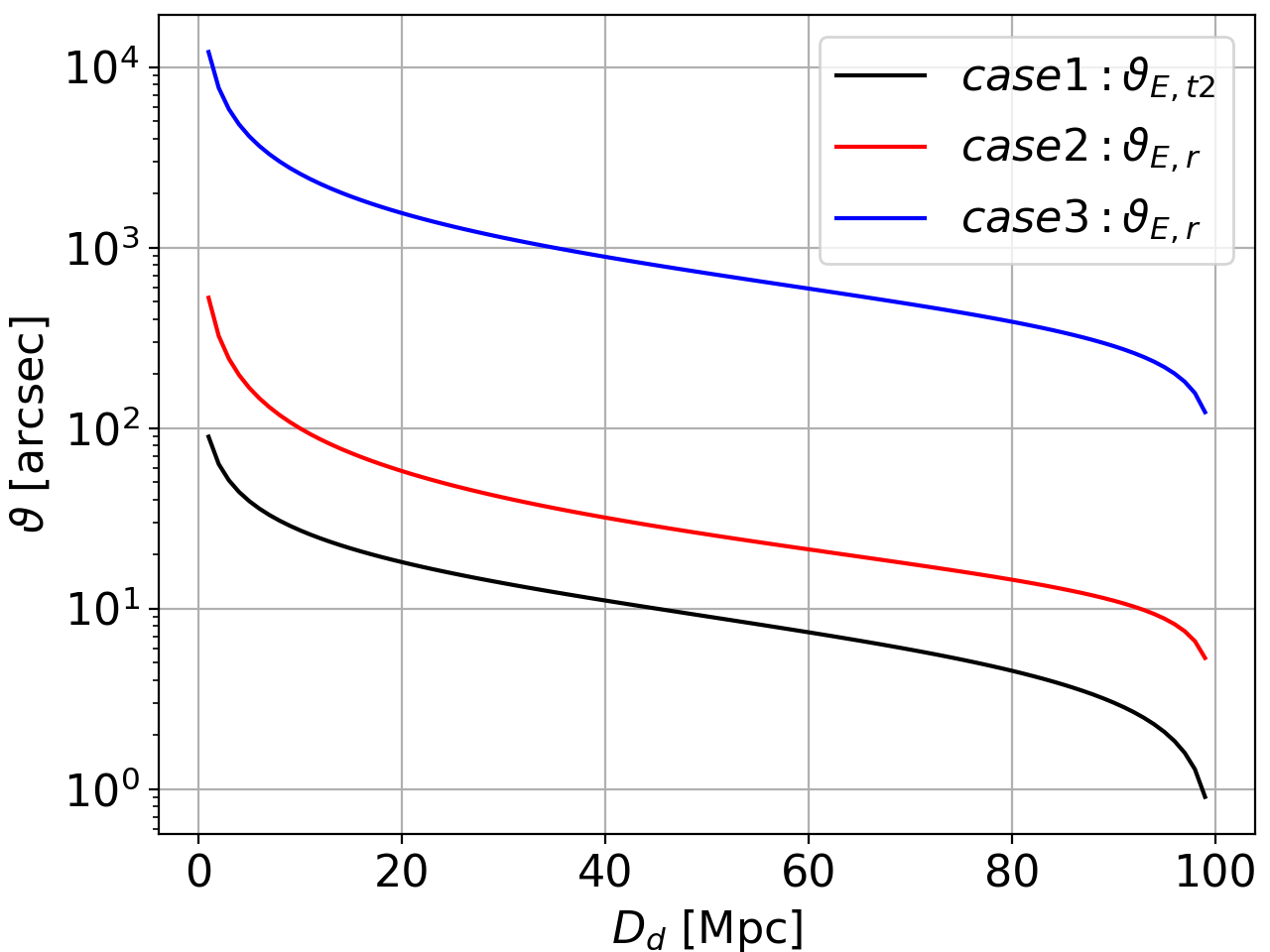}
\caption{Left: $\vartheta_\mathrm{E}$ of a SCAMP with fixed distance to us for different source distances in a $\Lambda$CDM-like cosmology. Right: $\vartheta_\mathrm{E}$ of a SCAMP at different distances between us and a source at $D_\mathrm{s}=100$~Mpc, i.~e.~in a least-cosmology-dependent lensing scenario. For case 1, three Einstein radii exist, shown in Fig.~\ref{fig:case1}. The largest one is plotted here, while for case~2 and 3, only radial Einstein radii exist.}
\label{fig:gravitational_results}
\end{figure*}

Fig.~\ref{fig:gravitational_results} (left) shows the results for the first scenario, Fig.~\ref{fig:gravitational_results} (right) the results for the second. 
For reference, the largest Einstein radius of $55$~arcsec has been observed for a galaxy cluster of mass $7.4\times 10^{14} \, M_\odot$ in \cite{bib:Zitrin}.
As can be read off Fig.~\ref{fig:gravitational_results}, this galaxy-cluster-scale Einstein radius is of the same order of magnitude as the Einstein radii of cases~1 to 2 for most distance ratio combinations.
Only the heaviest SCAMP considered here exceeds known sizes of strong gravitational lenses for all distance ratios.
Similar results are found in the scenario in our cosmic neighbourhood, apart from the limiting cases, forcing $\vartheta_\mathrm{E}$ to diverge for $D_\mathrm{d}$ towards zero and $\vartheta_{E}$ towards zero for $D_\mathrm{d}$ towards $D_\mathrm{s}$.

\begin{figure*}
\centering
\includegraphics[width=0.48\textwidth]{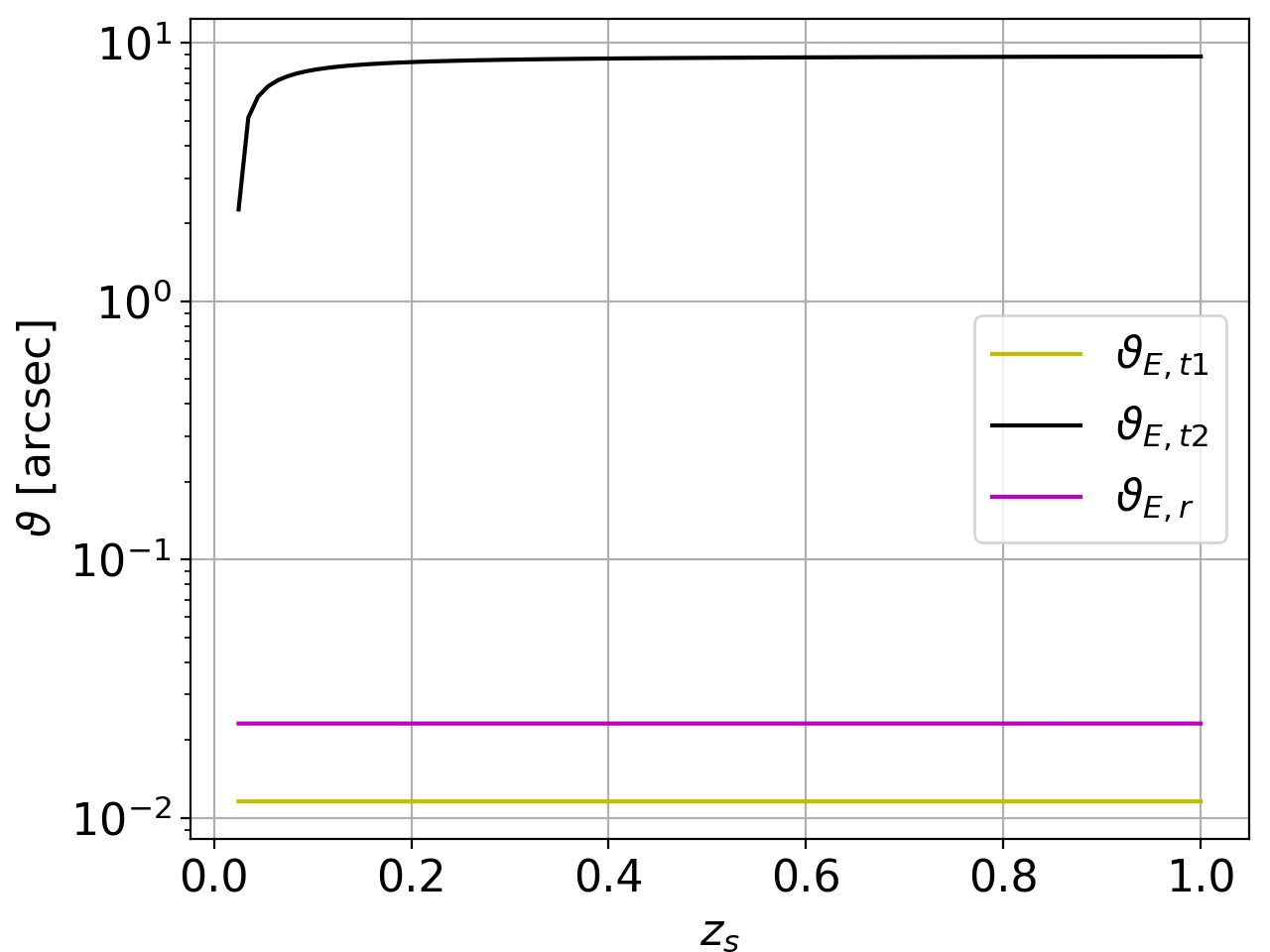}
\includegraphics[width=0.48\textwidth]{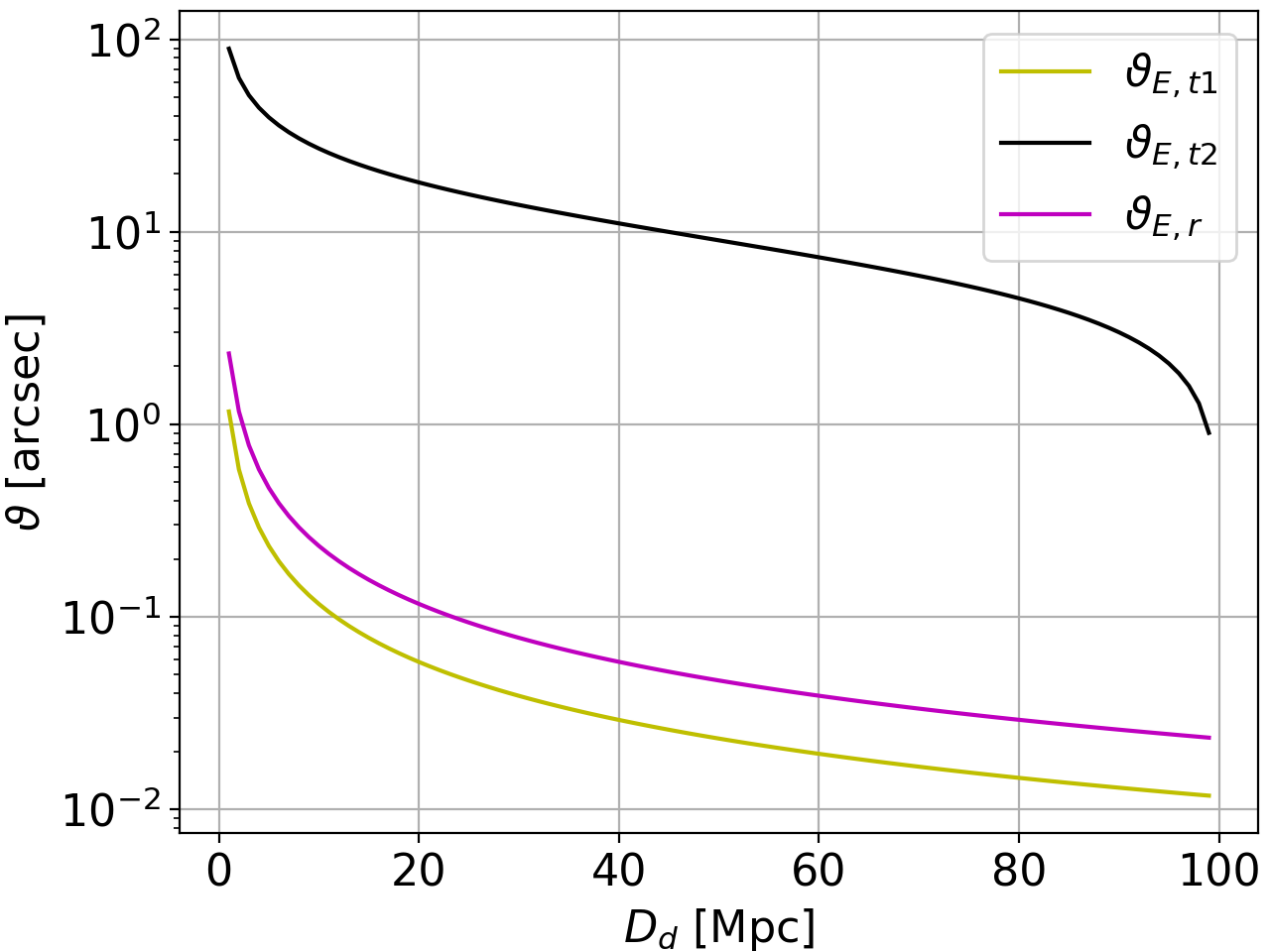}
\caption{Left: all $\vartheta_\mathrm{E}$ of a SCAMP of $10^{12} M_\odot$ with fixed $D_\mathrm{d}$ for different $D_\mathrm{s}$ in a $\Lambda$CDM-like cosmology. Right: all $\vartheta_\mathrm{E}$ of a SCAMP of $10^{12} M_\odot$ at different $D_\mathrm{d}$ and a source fixed at $D_\mathrm{s}=100$~Mpc, i.~e.~in a least-cosmology-dependent lensing scenario.}
\label{fig:case1}
\end{figure*}

One might think it could be possible to refute the existence of SCAMPs due to the extreme sizes of the Einstein radii for cases~2 and 3.
However, these Einstein radii are \emph{radial} $\vartheta_\mathrm{E}$, implying that they do not cause easily traceable Einstein rings but only radial arcs.
Moreover, cases 2 and 3 do not have any tangential critical curves due to their large $r_\mathrm{Q}/r_\mathrm{S}$ ratio. 
Only case 1 has three Einstein radii as shown for the fixed $D_\mathrm{d}$ and the fixed $D_\mathrm{s}$ in Fig.~\ref{fig:case1} (left) and (right), respectively.
With $\vartheta_\mathrm{E,t1} \approx 0.012''$ and $\vartheta_\mathrm{E,r} \approx 0.023''$ in Fig.~\ref{fig:case1} (left), the inner tangential and the radial critical curve are below the resolution limit of the space-based telescopes. 
For Fig.~\ref{fig:case1} (right), the same conclusion is reached for most lens distances. 
Only for $D_\mathrm{d} < 15$-20~Mpc, the inner critical curves may be spatially resolved. 
In addition to this requirement, the source must also be perfectly aligned behind the point-like singularity.
If this is the case, however, we receive a relatively unique signal which distinguishes a SCAMP from other weak-field gravitational lenses or Schwarzschild BHs with a single $\vartheta_\mathrm{E}$. 

%%%
%%%

%%%
%%%

Another good signature of SCAMPs is an observed galaxy-cluster-scale Einstein radius with a much smaller amount of luminous matter in its vicinity than usually expected in galaxy clusters. 
In the best case, SCAMPs occur outside of known luminous structures, as mentioned in Section~\ref{sec:obs_gravitational_field}. 
While radially symmetric critical curves of isolated SCAMPs can be easily distinguished from critical curves caused by ellipsoidal mass density profiles like galaxy clusters, future studies of SCAMPs as lenses will investigate whether external shear in their environment can also perturb their radially symmetric critical curves. 
It also remains a question for further studies whether the radial arcs for cases~2 and 3 can be distinguished from other elongated observables in the universe.
In any case, the evidence they provide in favour of SCAMPs is much weaker than large tangential Einstein radii, particularly because magnifications for the inner critical curves are much smaller than the ones for the outmost tangential one, \cite{bib:Virbhadra2002}.

To briefly comment on the strong lensing configurations being static, we assume a very fast relative motion between the background source and the SCAMP as a microlens of 1000~km/s, which is the order of velocities of galaxies moving in a galaxy-cluster-scale gravitational potential. 
If the SCAMP is at distance $D_\mathrm{d}=100$~Mpc to us, its Einstein radius inferred from Fig.~\ref{fig:gravitational_results} can be assumed to be 10~arcsec for some configurations. 
Then, the time to move a distance of one Einstein radius at $D_\mathrm{d}$ amounts to approximately $\delta t = 5 \times 10^{6}$~years.  
Even if we reduce the distance to $D_\mathrm{d}=1$~Mpc, the time to cross 10~arcsec is still $\delta t =5000$~years.

%%%%%%%%%%%%%%%%%%%%
\section{Conclusion}
\label{sec:conclusion}

Assuming that primordial extremely massive black holes (PEMBHs) of masses in the range of $10^{12}$ to $10^{14}~M_\odot$ actually exist, we explored the possible effects of these PEMBHs carrying the extreme charges as discussed in \cite{bib:Frampton3} to find suitable observables that could be searched for in sky surveys. 
As is known, any charged BH whose electric repulsion is larger than its gravitational attraction becomes a naked singularity, if it can be described as a Reissner-Nordstr\"{o}m (RN) BH.
Thus, the PEMBHs with the scaling between charge and mass according to \cite{bib:Frampton3} turn out to be naked singularities, which we called Stupenduously Charged And Massive Primordials (SCAMPs).
More precisely, the lack of a photonsphere even makes the SCAMPs \emph{strongly} naked singularities. 

Due to the lack of models for dynamical spacetimes containing two such SCAMPs mutually repelling each other, we restricted our analysis of observable signatures of SCAMPs to distances at which a Newtonian metric yields a good approximation to all possible effects, see Fig.~\ref{fig:approximation} for details. 
We thus modelled the SCAMPs as point charges being accelerated away from each other due to their net Coulomb repulsion. 

Subsequently, we determined the induced Li\'enard-Wiechert electro-magnetic fields for each SCAMP and found that all velocities and accelerations are well within the non-relativistic limit.
Travel times are shorter than the age of the universe, but much longer than human life times, such that we cannot observe the motion of these SCAMPs and changes in the electro-magnetic fields directly, see Table~\ref{tab:dynamics} for a summary.
The electro-magnetic fields at distances far from the SCAMPs, see Figs.~\ref{fig:E_tot} and \ref{fig:B_tot}, were found to be perturbing effects to the well-known emission and absorption spectra of neutral hydrogen for most configurations analysed.
In particular, these extreme sort of BH-singularities do not cause any proton decay, based on the proton decay model of \cite{bib:Wistisen2021}, or a dissociation of neutral hydrogen at distances where the Newtonian approximation is valid. 

Furthermore, we identified observations of rotation measures induced in plasma clouds close to a SCAMP as a promising signature.
Strong gravitational lensing deflecting background light into specific multiple-image configurations of high symmetry or Einstein rings with sizes that otherwise belong to galaxy-cluster-scale lenses are a second observable hint for the existence of such SCAMPs. 
For the strong-lensing signature, we used the strong-field lensing equations based on a strongly singular RN metric and investigated background sources in a $\Lambda$CDM-like cosmology, assuming that SCAMPs cause a background expansion equivalent to the one induced by dark energy. 
We also determined the observables for a light-deflecting SCAMP and a background source at distances out to 100~Mpc, because restricting the analysis to objects being in the linear Hubble flow at most allows for the least-cosmology-dependent search for individual SCAMPs. 

Hence, contrary to the first impression, such extreme structures containing strongly naked singularities could cause disruptive, catastrophic effects and immediately lead to the rejection of the hypothesis SCAMPs could replace dark energy, we showed that the far field of these extreme structures generates rather moderate perturbative phenomena. 
Similarly, the strong gravitational-field lensing effects that may leave a unique signature of two tangential Einstein radii for SCAMP masses below $10^{13} M_\odot$ require a sub-arcsecond spatial resolution and a precise fine-tuning in the alignment of observer, SCAMP, and background light source. 
For SCAMP masses of $10^{13} M_\odot$ and more, only radial Einstein arcs can occur as the charge-to-mass ratio of the strongly naked singularity impedes the generation of tangential critical curves. 

Nevertheless, the particular imprints in observables like rotation measures and strong gravitational lensing effects may be detected in sky surveys in radio and optical wave bands with current and upcoming telescopes. 
It is also possible that existing observations found to challenge the cosmological principle as summarised in \cite{bib:Aluri2022} already contain signatures of SCAMPs.  
A forecast on the detailed probabilities to find the signatures developed in Section~\ref{sec:observable_signatures} is left to future work, as is the development of statistical observables for a cosmic SCAMP ensemble.

%%%%%%%%%%%%%%%%%%%%%%%%%%%%%%%%%%%
\appendix
\section{Derivation of the non-relativistic Li\'enard-Wiechert fields}
\label{app:derivation}
The general formula for the electric field of a moving source with charge $q$ is given by
\begin{align}
\boldsymbol{E}(\boldsymbol{D}, T) = \dfrac{1}{4\pi\epsilon_0} &\left( \dfrac{ q \left( \boldsymbol{e}_D - \boldsymbol{\beta}_\mathrm{s}(t) \right) }{\gamma^2 \left(1 - \boldsymbol{e}_\mathrm{D} \cdot \boldsymbol{\beta}_\mathrm{s} \right)^3 \left| \boldsymbol{D} \right|^2} \; + \dfrac{q \, \boldsymbol{e}_{D} \times \left( \left( \boldsymbol{e}_{D} - \boldsymbol{\beta}_\mathrm{s} \right) \times \dot{\boldsymbol{\beta}}_\mathrm{s} \right)}{\left(1 - \boldsymbol{e}_{D} \cdot \boldsymbol{\beta}_\mathrm{s} \right)^3 \left| \boldsymbol{D} \right|} \right)_{t} \;,
\label{app_eq:E1}
\end{align}
where $\boldsymbol{\beta}_\mathrm{s} \equiv \boldsymbol{v}_\mathrm{s}/c$, $\gamma = 1/\sqrt{1-|\boldsymbol{\beta}|^2}$ determine the relativistic motion of the source in direction $\boldsymbol{e}_\mathrm{s}$, $\dot{\boldsymbol{\beta}}_\mathrm{s}$ denotes its acceleration, and the source and the observer are separated by a distance $\boldsymbol{D}$ with direction $\boldsymbol{e}_D$. 
The term in large brackets is evaluated at the retarded time $t$, as seen by the observer. 
In the case of the moving black hole introduced in Section~\ref{sec:electro-magnetic_fields}, the retarded time is $\delta t$ when the black hole has moved $2r_1$ in $x$-direction. 
The motion is non-relativistic, such that $\boldsymbol{\beta}_\mathrm{s} \approx 0$ (implying $\gamma \approx 1$).
Hence, \eqref{app_eq:E1} can be simplified to
\begin{equation}
\boldsymbol{E}(\boldsymbol{D}, T) = \dfrac{1}{4\pi \epsilon_0} \left( \dfrac{ q\, \boldsymbol{e}_D}{\left| \boldsymbol{D} \right|^2} + \dfrac{q \, \boldsymbol{e}_{D} \times\left( \boldsymbol{e}_{D} \times \dot{\boldsymbol{\beta}}_\mathrm{s} \right)}{\left| \boldsymbol{D} \right|} \right)_{t} \;.
\label{app_eq:E2}
\end{equation}
Next, we use $\boldsymbol{e}_D \cdot \boldsymbol{e}_D = 1$, and 
\begin{equation}
\boldsymbol{e}_D \times \left( \boldsymbol{e}_D \times \dot{\boldsymbol{\beta}}_\mathrm{s} \right) = \boldsymbol{e}_D \left( \dot{\boldsymbol{\beta}}_\mathrm{s}  \cdot \boldsymbol{e}_D \right) - \dot{\boldsymbol{\beta}}_\mathrm{s} \left(\boldsymbol{e}_D \cdot \boldsymbol{e}_D \right)
\label{app_eq:cross_product}
\end{equation}
to simplify the last term on the right-hand side to arrive at \eqref{eq:E}. 

Analogously, in the same notation, the general formula for the magnetic field caused by a moving source with charge $q$ is given by
\begin{align}
\boldsymbol{B}(\boldsymbol{D}, T) = \dfrac{\mu_0}{4\pi} &\left( \dfrac{ q c \,  \boldsymbol{\beta}_\mathrm{s}(t) \times \boldsymbol{e}_D}{\gamma^2 \left(1 - \boldsymbol{e}_\mathrm{D} \cdot \boldsymbol{\beta}_\mathrm{s} \right)^3 \left| \boldsymbol{D} \right|^2} \; +  \dfrac{q \, \boldsymbol{e}_{D} \times \left( \boldsymbol{e}_{D} \times \left( \left( \boldsymbol{e}_{D} - \boldsymbol{\beta}_\mathrm{s} \right) \times \dot{\boldsymbol{\beta}}_\mathrm{s} \right) \right)}{\left(1 - \boldsymbol{e}_{D} \cdot \boldsymbol{\beta}_\mathrm{s} \right)^3 \left| \boldsymbol{D} \right|} \right)_{t} \;.
\label{app_eq:B1}
\end{align}
Constraining \eqref{app_eq:B1} to non-relativistic cases, we obtain
\begin{align}
\boldsymbol{B}(\boldsymbol{D}, T) = \dfrac{\mu_0}{4\pi} &\left( \dfrac{ q\, \boldsymbol{v}_\mathrm{s}(t) \times \boldsymbol{e}_D}{\left| \boldsymbol{D} \right|^2} \; +  \dfrac{q \, \boldsymbol{e}_{D} \times \left( \boldsymbol{e}_{D} \times \left( \boldsymbol{e}_{D} \times \dot{\boldsymbol{\beta}}_\mathrm{s} \right) \right)}{\left| \boldsymbol{D} \right|} \right)_{t} \;.
\label{app_eq:B2}
\end{align}
Next, we use that $\boldsymbol{e}_D \times \boldsymbol{e}_D = 0$, $\boldsymbol{e}_D \cdot \boldsymbol{e}_D = 1$, and \eqref{app_eq:cross_product} to simplify the last term on the right-hand side to
\begin{equation}
\boldsymbol{B}(\boldsymbol{D}, T) = \dfrac{\mu_0}{4\pi} \left( \dfrac{ q\, \boldsymbol{v}_\mathrm{s}(t) \times \boldsymbol{e}_D}{\left| \boldsymbol{D} \right|^2} - \dfrac{q \, \boldsymbol{e}_{D} \times \dot{\boldsymbol{\beta}}_\mathrm{s}}{\left| \boldsymbol{D} \right|} \right)_{t} \;.
\label{app_eq:B3}
\end{equation}
With $\boldsymbol{e}_{D} \times \dot{\boldsymbol{\beta}}_\mathrm{s} = - \dot{\boldsymbol{\beta}}_\mathrm{s} \times \boldsymbol{e}_{D}$ and dissecting the dynamical properties of the source into their amplitude and the direction $\boldsymbol{e}_\mathrm{s}$, i.~e.~the $x$-direction in our case, we arrive at \eqref{eq:B}.

%%%%%%%%%%%%%%%%%%%%%%%%%
\section{Gravitational lensing in RN spacetimes}
\label{app:lensing}

\begin{figure}
\centering
\includegraphics[width=0.65\textwidth]{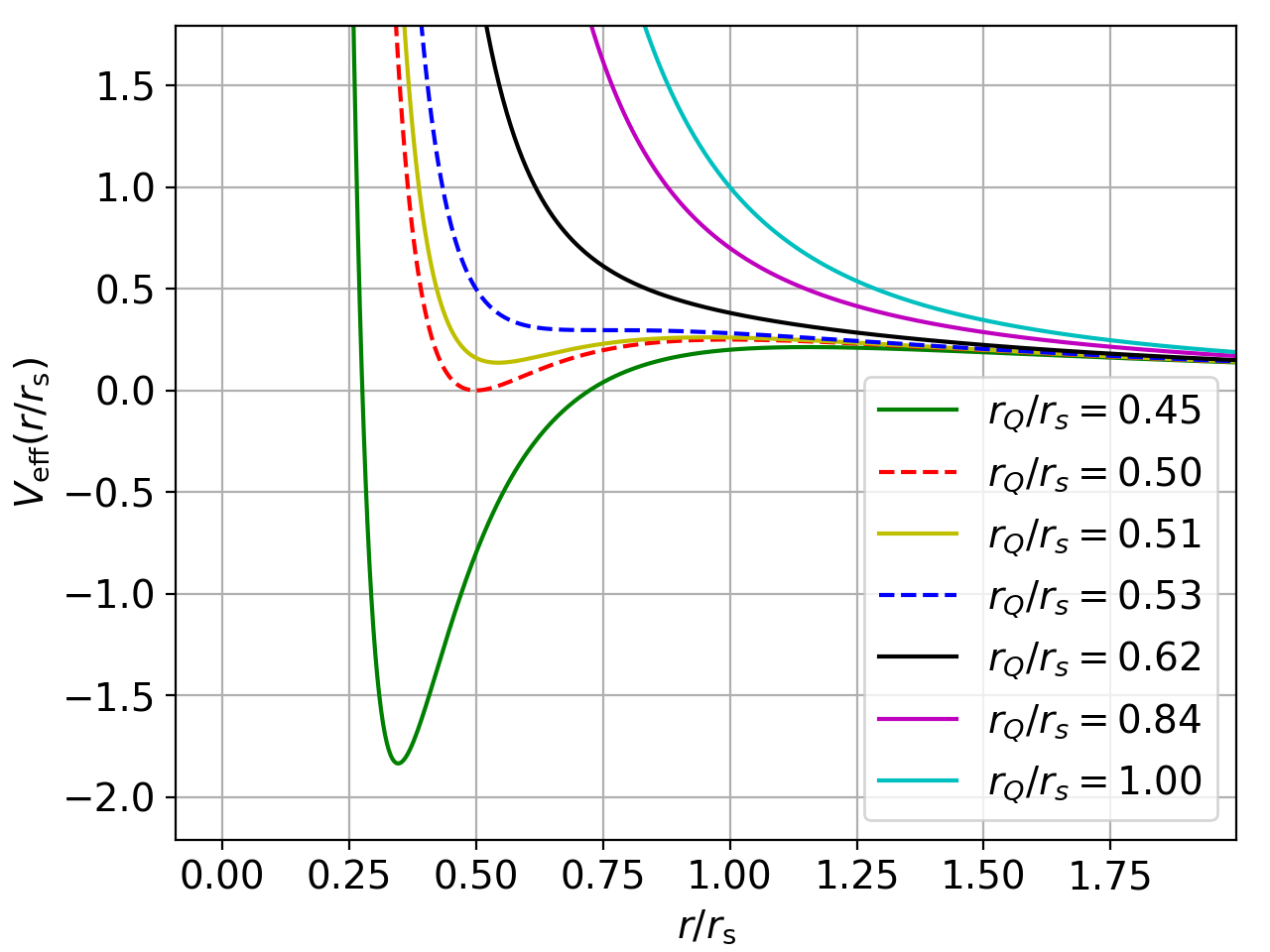}
\caption{Summary of possible geometries for null geodesics according to \eqref{eq:drdphi}: effective potential, \eqref{eq:Veff}, with respect to the dimensionless radius $x=r/r_\mathrm{S}$ for a RN BH with two event horizons and photon spheres ($r_\mathrm{Q}/r_\mathrm{S}=0.45$, green line), for the limiting case of an extremal BH ($r_\mathrm{Q}/r_\mathrm{S}=0.5$, red dashed line), for a weakly naked singularity without an event horizon ($r_\mathrm{Q}/r_\mathrm{S}=0.51$, yellow line), a singularity at the verge of becoming a strongly naked one ($r_\mathrm{Q}/r_\mathrm{S}=0.53$, blue dashed line), and three further strongly naked singularities with increasing ratio ($r_\mathrm{Q}/r_\mathrm{S}\ge 0.62$, black, magenta, cyan lines). Depending on the impact parameter, bound or unbound orbits occur for different $r_\mathrm{b}^{-2}$ in the RN BH and weakly naked singularity metrics. For strongly naked singularities, there are only scattering, unbound orbits. For comparison, the SCAMPs in Section~\ref{sec:application_examples} have $r_\mathrm{Q}/r_\mathrm{S} > 7$.}
\label{fig:RN_geometry}
\end{figure}

Following \cite{bib:Hackmann2008}, the null geodesic determining the trajectory of light in the RN metric of \eqref{eq:RN} is assumed to have $\vartheta = \pi/2$ without loss of generality, such that it can be described by
\begin{equation}
\left( \dfrac{{\rm d}r}{{\rm d} \varphi}\right)^2 = r^4 \left( \dfrac{1}{r_\mathrm{b}^2} - \dfrac{1}{r^2} \left(1 - \dfrac{r_\mathrm{S}}{r} + \dfrac{r_\mathrm{Q}^2}{r^2}\right) \right) \;.
\label{eq:drdphi}
\end{equation}
Abbreviating
\begin{equation}
V_\mathrm{eff}(r) \equiv \dfrac{1}{r^2} \left(1 - \dfrac{r_\mathrm{S}}{r} + \dfrac{r_\mathrm{Q}^2}{r^2}\right) \;,
\label{eq:Veff}
\end{equation}
the radii of the event horizons, if they exist, are determined as the roots of $V_\mathrm{eff}(r)$. 
The radii of the photon spheres are determined by the extrema of $V_\mathrm{eff}(r)$. 
Fig.~\ref{fig:RN_geometry} summarises the possible geometries for representative ratios of $r_\mathrm{Q}/r_\mathrm{S}$, showing a RN BH with two event horizons and photon spheres ($r_\mathrm{Q}/r_\mathrm{S}=0.45$), the limiting case of an extremal BH ($r_\mathrm{Q}/r_\mathrm{S}=0.5$), a weakly naked singularity without an event horizon ($r_\mathrm{Q}/r_\mathrm{S}=0.51$), a singularity at the verge of becoming a strongly naked one ($r_\mathrm{Q}/r_\mathrm{S}=0.53$), and three further strongly naked singularities with increasing ratio ($r_\mathrm{Q}/r_\mathrm{S}\ge 0.62$). 
Having fixed the geometry of all possible light trajectories in this way, the path taken depends on the impact parameter $r_\mathrm{b}$.

%%%
%%%

\begin{figure}
\centering
\includegraphics[width=0.6\textwidth]{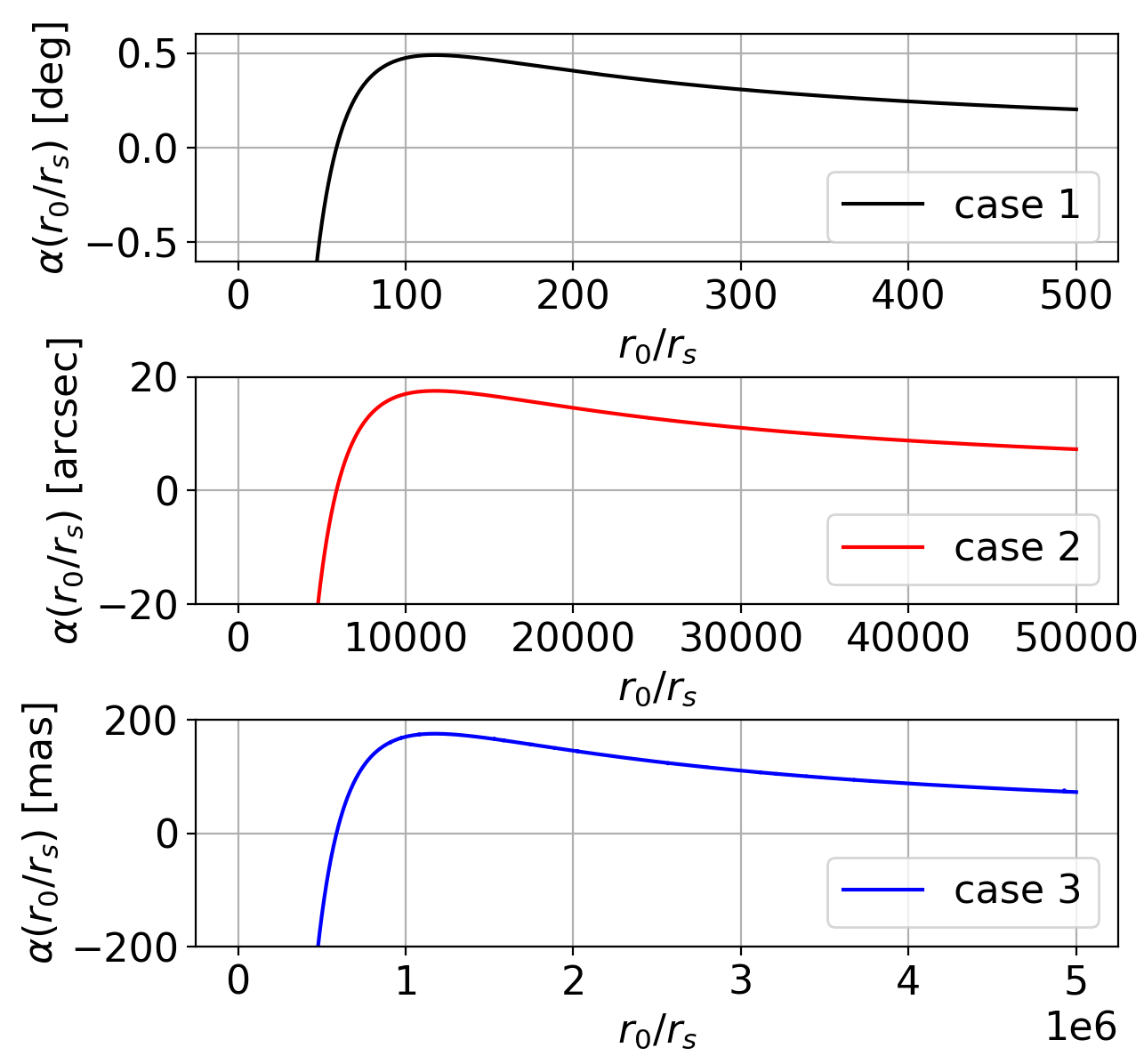}
\caption{Deflection angles $\alpha(r_0/r_\mathrm{S})$ dependent on the radius of closest approach $r_0$ given by \eqref{eq:alpha}: for $M=10^{12} M_\odot$ (top), $M=10^{13} M_\odot$ (centre), and $M=10^{14} M_\odot$ (bottom).}
\label{fig:deflection_angle_RN}
\end{figure}

Since \eqref{eq:alpha} is numerically unstable for $x_\mathrm{Q} \gg x_\mathrm{S}$, we substitute $u \equiv 1/x$ to obtain 
\begin{equation}
\alpha(u_0) = 2 \int \limits_0^{u_0} \dfrac{{\rm{d}}u}{\sqrt{u_\mathrm{b}^2 - u^2 \left( 1 - u + x_\mathrm{Q}^2 u^2 \right)}} \;,
\end{equation}
with $u_\mathrm{b} = 1/x_\mathrm{b}$ using $x_\mathrm{b}$ as defined in \eqref{eq:impact}. 
For the strongly naked singularities considered in this paper, $x_0$ is the only real, positive root of the denominator.
The second root is negative, the third and fourth root are complex. 
To show that the SCAMP deflection angles based on a RN metric are similar to those of the Janis-Newman-Winicour naked singularities discussed in \cite{bib:Virbhadra2002}, Fig.~\ref{fig:deflection_angle_RN} shows $\alpha(r_0/r_\mathrm{s})$ for the RN naked singularities introduced in Section~\ref{sec:application_examples}.

%%%
%%%

\begin{figure}
\centering
\includegraphics[width=0.6\textwidth]{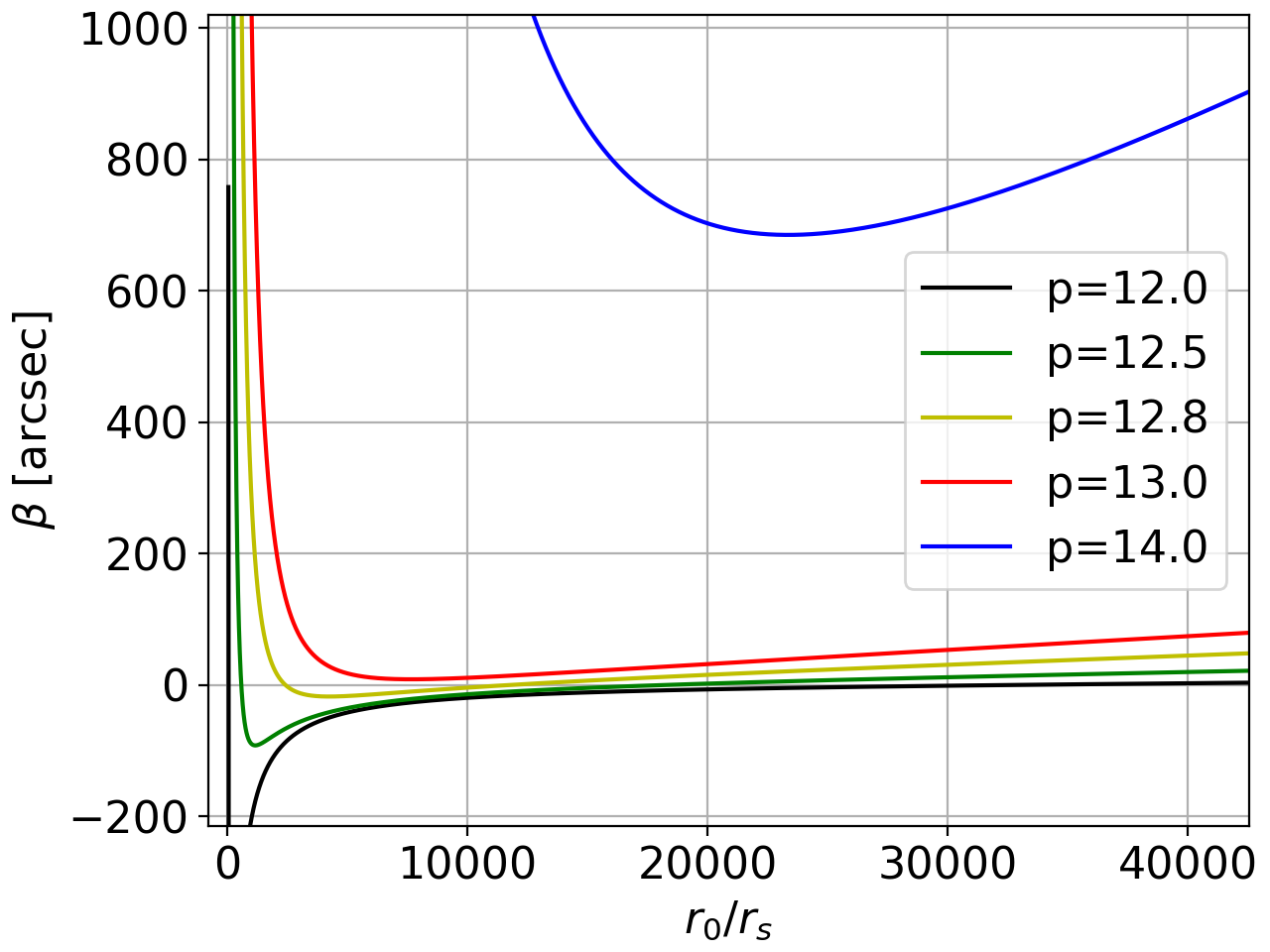}
\caption{Lens equation ${\rm tan}(\beta) \approx \beta$ for different SCAMPs at $z_\mathrm{d}=0.024$ with $z_\mathrm{s}=0.05$ to illustrate the change in the number of critical curves. Below $10^{13}~M_\odot$, a SCAMP has two tangential Einstein radii (zero crossings of the function) and a radial one (minimum of the function), while above, only a radial Einstein radius remains.}
\label{fig:tanbeta}
\end{figure}

Inserting this deflection angle into the strong-field lens equation, \eqref{eq:strong_le}, the zero crossings of the lens equation determine the tangential Einstein radii, while the position of the minimum yields the radial Einstein radius. 
As an example, Fig.~\ref{fig:tanbeta} shows the relevant detail for small values $x_0$ and correspondingly small values of $\beta$, assuming a SCAMP at $z_\mathrm{d}=0.024$ and a background source at $z_\mathrm{s}=0.05$, such that the small-angle approximation holds. 
From this plot, it becomes clear that the geometry of the critical curves changes for SCAMPs of about $10^{13} M_\odot$. 
The $r_\mathrm{Q}/r_\mathrm{S}$ ratio becomes so large that the two tangential Einstein radii disappear and only a radial one remains. 

%%%
%%%

\acknowledgments

I would like to thank Paul Frampton, Eduardo Guendelman, Thomas L.~Curtright, and the entire BASIC2022 workshop for the inspiring discussions raising my interest in this topic. Moreover, I am grateful for further helpful comments and discussion with Leonid Gurvits, Kumar Shwetketu Virbhadra, Avi Loeb, and David Benisty. 
Thanks a lot in addition to R\"{u}diger Vaas for very helpful comments on the paper draft.

%\paragraph{Note added.} This is also a good position for notes added after the paper has been written.


\bibliographystyle{jhep}
\bibliography{charged_bh} % if your bibtex file is called example.bib

\providecommand{\href}[2]{#2}\begingroup\raggedright\begin{thebibliography}{10}

\bibitem{bib:Hawking1971}
S.~{Hawking}, \emph{{Gravitationally collapsed objects of very low mass}},
  \href{https://doi.org/10.1093/mnras/152.1.75}{\emph{Monthly Notices of the
  Royal Astronomical Society} {\bfseries 152} (1971) 75}.

\bibitem{bib:Zeldovic1966}
Y.B.~{Zel'dovich} and I.D.~{Novikov}, \emph{{The Hypothesis of Cores Retarded
  during Expansion and the Hot Cosmological Model}}, {\emph{Astronomicheskii
  Zhurnal} {\bfseries 43} (1966) 758}.

\bibitem{bib:Cappelluti2022}
N.~{Cappelluti}, G.~{Hasinger} and P.~{Natarajan}, \emph{{Exploring the
  High-redshift PBH-{\ensuremath{\Lambda}}CDM Universe: Early Black Hole
  Seeding, the First Stars and Cosmic Radiation Backgrounds}},
  \href{https://doi.org/10.3847/1538-4357/ac332d}{\emph{The Astrophysical
  Journal} {\bfseries 926} (2022) 205}
  [\href{https://arxiv.org/abs/2109.08701}{{\ttfamily 2109.08701}}].

\bibitem{bib:Carr2022}
B.~{Carr} and F.~{K{\"u}hnel}, \emph{{Primordial black holes as dark matter
  candidates}},
  \href{https://doi.org/10.21468/SciPostPhysLectNotes.48}{\emph{SciPost Phys.
  Lect. Notes} (2022) 48}.

\bibitem{bib:Villanueva2021}
P.~{Villanueva-Domingo}, O.~{Mena} and S.~{Palomares-Ruiz}, \emph{{A brief
  review on primordial black holes as dark matter}},
  \href{https://doi.org/10.3389/fspas.2021.681084}{\emph{Frontiers in Astronomy
  and Space Sciences} {\bfseries 8} (2021) 87}
  [\href{https://arxiv.org/abs/2103.12087}{{\ttfamily 2103.12087}}].

\bibitem{bib:Carr_stup}
B.~{Carr}, F.~{K{\"u}hnel} and L.~{Visinelli}, \emph{{Constraints on
  stupendously large black holes}},
  \href{https://doi.org/10.1093/mnras/staa3651}{\emph{Monthly Notices of the
  Royal Astronomical Society} {\bfseries 501} (2021) 2029}
  [\href{https://arxiv.org/abs/2008.08077}{{\ttfamily 2008.08077}}].

\bibitem{bib:Boylan2023}
M.~{Boylan-Kolchin}, \emph{{Stress testing {\ensuremath{\Lambda}}CDM with
  high-redshift galaxy candidates}},
  \href{https://doi.org/10.1038/s41550-023-01937-7}{\emph{Nature Astronomy}
  {\bfseries 7} (2023) 731} [\href{https://arxiv.org/abs/2208.01611}{{\ttfamily
  2208.01611}}].

\bibitem{bib:Kocevski2023}
D.D.~{Kocevski}, M.~{Onoue}, K.~{Inayoshi}, J.R.~{Trump}, P.~{Arrabal Haro},
  A.~{Grazian} et~al., \emph{{Hidden Little Monsters: Spectroscopic
  Identification of Low-mass, Broad-line AGNs at $z > 5$ with CEERS}},
  \href{https://doi.org/10.3847/2041-8213/ace5a0}{\emph{The Astrophysical
  Journal Letters} {\bfseries 954} (2023) L4}
  [\href{https://arxiv.org/abs/2302.00012}{{\ttfamily 2302.00012}}].

\bibitem{bib:Labbe2023}
I.~{Labbe}, J.E.~{Greene}, R.~{Bezanson}, S.~{Fujimoto}, L.J.~{Furtak},
  A.D.~{Goulding} et~al., \emph{{UNCOVER: Candidate Red Active Galactic Nuclei
  at 3<z<7 with JWST and ALMA}},
  \href{https://doi.org/10.48550/arXiv.2306.07320}{\emph{arXiv e-prints} (2023)
  arXiv:2306.07320} [\href{https://arxiv.org/abs/2306.07320}{{\ttfamily
  2306.07320}}].

\bibitem{bib:Matthee2023}
J.~{Matthee}, R.P.~{Naidu}, G.~{Brammer}, J.~{Chisholm}, A.-C.~{Eilers},
  A.~{Goulding} et~al., \emph{{Little Red Dots: an abundant population of faint
  AGN at $z\sim5$ revealed by the EIGER and FRESCO JWST surveys}},
  \href{https://doi.org/10.48550/arXiv.2306.05448}{\emph{arXiv e-prints} (2023)
  arXiv:2306.05448} [\href{https://arxiv.org/abs/2306.05448}{{\ttfamily
  2306.05448}}].

\bibitem{bib:Pacucci2023}
F.~{Pacucci}, B.~{Nguyen}, S.~{Carniani}, R.~{Maiolino} and X.~{Fan},
  \emph{{JWST CEERS and JADES Active Galaxies at z = 4-7 Violate the Local M
  $_{{\textbullet}}$-M $_{{\ensuremath{\star}}}$ Relation at
  >3{\ensuremath{\sigma}}: Implications for Low-mass Black Holes and Seeding
  Models}}, \href{https://doi.org/10.3847/2041-8213/ad0158}{\emph{The
  Astrophysical Journal Letters} {\bfseries 957} (2023) L3}
  [\href{https://arxiv.org/abs/2308.12331}{{\ttfamily 2308.12331}}].

\bibitem{bib:Carr2021}
B.~{Carr}, K.~{Kohri}, Y.~{Sendouda} and J.~{Yokoyama}, \emph{{Constraints on
  primordial black holes}},
  \href{https://doi.org/10.1088/1361-6633/ac1e31}{\emph{Reports on Progress in
  Physics} {\bfseries 84} (2021) 116902}
  [\href{https://arxiv.org/abs/2002.12778}{{\ttfamily 2002.12778}}].

\bibitem{bib:Frampton2}
P.H.~{Frampton}, \emph{{Possibility of Additional Intergalactic and
  Cosmological Dark Matter}}, {\emph{arXiv e-prints} (2022) arXiv:2207.12408}
  [\href{https://arxiv.org/abs/2207.12408}{{\ttfamily 2207.12408}}].

\bibitem{bib:Frampton1}
P.H.~{Frampton}, \emph{{Entropy of the Universe and Hierarchical Dark Matter}},
  \href{https://doi.org/10.3390/e24081171}{\emph{Entropy} {\bfseries 24} (2022)
  1171} [\href{https://arxiv.org/abs/2202.04432}{{\ttfamily 2202.04432}}].

\bibitem{bib:Planck2020}
{Planck Collaboration}, \emph{{Planck 2018 results. VI. Cosmological
  parameters}},
  \href{https://doi.org/10.1051/0004-6361/201833910}{\emph{Astronomy \&
  Astrophysics} {\bfseries 641} (2020) A6}
  [\href{https://arxiv.org/abs/1807.06209}{{\ttfamily 1807.06209}}].

\bibitem{bib:Farrah2023}
D.~{Farrah}, K.S.~{Croker}, M.~{Zevin}, G.~{Tarl{\'e}}, V.~{Faraoni},
  S.~{Petty} et~al., \emph{{Observational Evidence for Cosmological Coupling of
  Black Holes and its Implications for an Astrophysical Source of Dark
  Energy}}, \href{https://doi.org/10.3847/2041-8213/acb704}{\emph{The
  Astrophysical Journal Letters} {\bfseries 944} (2023) L31}
  [\href{https://arxiv.org/abs/2302.07878}{{\ttfamily 2302.07878}}].

\bibitem{bib:Frampton3}
P.H.~{Frampton}, \emph{{Electromagnetic accelerating universe}},
  \href{https://doi.org/10.1016/j.physletb.2022.137480}{\emph{Physics Letters
  B} {\bfseries 835} (2022) 137480}
  [\href{https://arxiv.org/abs/2210.10632}{{\ttfamily 2210.10632}}].

\bibitem{bib:Frampton4}
P.H.~{Frampton}, \emph{{A model of dark matter and energy}},
  \href{https://doi.org/10.1142/S0217732323500323}{\emph{Modern Physics Letters
  A} {\bfseries 38} (2023) 2350032}
  [\href{https://arxiv.org/abs/2301.10719}{{\ttfamily 2301.10719}}].

\bibitem{bib:Croker2019}
K.S.~{Croker} and J.L.~{Weiner}, \emph{{Implications of Symmetry and Pressure
  in Friedmann Cosmology. I. Formalism}},
  \href{https://doi.org/10.3847/1538-4357/ab32da}{\emph{The Astrophysical
  Journal} {\bfseries 882} (2019) 19}
  [\href{https://arxiv.org/abs/2107.06643}{{\ttfamily 2107.06643}}].

\bibitem{bib:Mistele2023}
T.~{Mistele}, \emph{{Comment on ``Observational Evidence for Cosmological
  Coupling of Black Holes and its Implications for an Astrophysical Source of
  Dark Energy''}},
  \href{https://doi.org/10.3847/2515-5172/acd767}{\emph{Research Notes of the
  American Astronomical Society} {\bfseries 7} (2023) 101}
  [\href{https://arxiv.org/abs/2304.09817}{{\ttfamily 2304.09817}}].

\bibitem{bib:Gaur2023}
R.~{Gaur} and M.~{Visser}, \emph{{Black holes embedded in FLRW cosmologies}},
  \href{https://doi.org/10.48550/arXiv.2308.07374}{\emph{arXiv e-prints} (2023)
  arXiv:2308.07374} [\href{https://arxiv.org/abs/2308.07374}{{\ttfamily
  2308.07374}}].

\bibitem{bib:Araya2022}
I.J.~{Araya}, N.D.~{Padilla}, M.E.~{Rubio}, J.~{Sureda}, J.~{Maga{\~n}a} and
  L.~{Osorio}, \emph{{Dark matter from primordial black holes would hold
  charge}}, \href{https://doi.org/10.1088/1475-7516/2023/02/030}{\emph{Journal
  of Cosmology and Astroparticle Physics} {\bfseries 2023} (2023) 030}
  [\href{https://arxiv.org/abs/2207.05829}{{\ttfamily 2207.05829}}].

\bibitem{bib:Carter1968}
B.~{Carter}, \emph{{Global structure of the Kerr family of gravitational
  fields}}, \href{https://doi.org/10.1103/PhysRev.174.1559}{\emph{Physical
  Review} {\bfseries 174} (1968) 1559}.

\bibitem{bib:Burrinskii2008}
A.~{Burinskii}, \emph{{The Dirac-Kerr-Newman electron}},
  \href{https://doi.org/10.1134/S0202289308020011}{\emph{Gravitation and
  Cosmology} {\bfseries 14} (2008) 109}
  [\href{https://arxiv.org/abs/hep-th/0507109}{{\ttfamily hep-th/0507109}}].

\bibitem{bib:Wald1974}
R.~{Wald}, \emph{{Gedanken experiments to destroy a black hole.}},
  \href{https://doi.org/10.1016/0003-4916(74)90125-0}{\emph{Annals of Physics}
  {\bfseries 82} (1974) 548}.

\bibitem{bib:Morris2023}
J.R.~{Morris}, \emph{{Effects of a modified Reissner-Nordstr{\"o}m spacetime}},
  \href{https://doi.org/10.48550/arXiv.2311.10890}{\emph{arXiv e-prints} (2023)
  arXiv:2311.10890} [\href{https://arxiv.org/abs/2311.10890}{{\ttfamily
  2311.10890}}].

\bibitem{bib:Hod2018}
S.~{Hod}, \emph{{Black-hole evaporation, cosmic censorship, and a quantum lower
  bound on the Bekenstein-Hawking temperature}},
  \href{https://doi.org/10.1140/epjc/s10052-018-6128-y}{\emph{European Physical
  Journal C} {\bfseries 78} (2018) 634}
  [\href{https://arxiv.org/abs/1809.04612}{{\ttfamily 1809.04612}}].

\bibitem{bib:Penrose1998}
R.~{Penrose}, \emph{{The Question of Cosmic Censorship}},  in \emph{Black Holes
  and Relativistic Stars}, R.M.~{Wald}, ed., p.~103, Jan., 1998.

\bibitem{bib:Singh1999}
T.P.~{Singh}, \emph{{Gravitational Collapse, Black Holes and Naked
  Singularities}}, \href{https://doi.org/10.1007/BF02702354}{\emph{Journal of
  Astrophysics and Astronomy} {\bfseries 20} (1999) 221}
  [\href{https://arxiv.org/abs/gr-qc/9805066}{{\ttfamily gr-qc/9805066}}].

\bibitem{bib:Sahu2012}
S.~{Sahu}, M.~{Patil}, D.~{Narasimha} and P.S.~{Joshi}, \emph{{Can strong
  gravitational lensing distinguish naked singularities from black holes?}},
  \href{https://doi.org/10.1103/PhysRevD.86.063010}{\emph{Physical Review D}
  {\bfseries 86} (2012) 063010}
  [\href{https://arxiv.org/abs/1206.3077}{{\ttfamily 1206.3077}}].

\bibitem{bib:Mummery2023}
A.~{Mummery}, S.~{Balbus} and A.~{Ingram}, \emph{{Testing theories of accretion
  and gravity with super-extremal Kerr discs}}, {\emph{arXiv e-prints} (2023)
  arXiv:2311.15742} [\href{https://arxiv.org/abs/2311.15742}{{\ttfamily
  2311.15742}}].

\bibitem{bib:Virbhadra2002}
K.S.~{Virbhadra} and G.F.~{Ellis}, \emph{{Gravitational lensing by naked
  singularities}},
  \href{https://doi.org/10.1103/PhysRevD.65.103004}{\emph{Physical Review D}
  {\bfseries 65} (2002) 103004}.

\bibitem{bib:Tsukamoto2021}
N.~{Tsukamoto}, \emph{{Gravitational lensing by a photon sphere in a
  Reissner-Nordstr{\"o}m naked singularity spacetime in strong deflection
  limits}}, \href{https://doi.org/10.1103/PhysRevD.104.124016}{\emph{Physical
  Review D} {\bfseries 104} (2021) 124016}.

\bibitem{bib:Sorkin2001}
E.~{Sorkin} and T.~{Piran}, \emph{{Formation and evaporation of charged black
  holes}}, \href{https://doi.org/10.1103/PhysRevD.63.124024}{\emph{Physical
  Review D} {\bfseries 63} (2001) 124024}
  [\href{https://arxiv.org/abs/gr-qc/0103090}{{\ttfamily gr-qc/0103090}}].

\bibitem{bib:Bozzola2021}
G.~{Bozzola} and V.~{Paschalidis}, \emph{{Numerical-relativity simulations of
  the quasicircular inspiral and merger of nonspinning, charged black holes:
  Methods and comparison with approximate approaches}},
  \href{https://doi.org/10.1103/PhysRevD.104.044004}{\emph{Physical Review D}
  {\bfseries 104} (2021) 044004}
  [\href{https://arxiv.org/abs/2104.06978}{{\ttfamily 2104.06978}}].

\bibitem{bib:Pesce2019}
D.~{Pesce}, K.~{Haworth}, G.J.~{Melnick}, L.~{Blackburn}, M.~{Wielgus},
  M.D.~{Johnson} et~al., \emph{{Extremely long baseline interferometry with
  Origins Space Telescope}},  in \emph{Bulletin of the American Astronomical
  Society}, vol.~51, p.~176, Sept., 2019,
  \href{https://doi.org/10.48550/arXiv.1909.01408}{DOI}
  [\href{https://arxiv.org/abs/1909.01408}{{\ttfamily 1909.01408}}].

\bibitem{bib:Novikov2021}
I.D.~{Novikov}, S.F.~{Likhachev}, Y.A.~{Shchekinov}, A.S.~{Andrianov},
  A.M.~{Baryshev}, A.I.~{Vasyunin} et~al., \emph{{Objectives of the Millimetron
  Space Observatory science program and technical capabilities of its
  realization}},
  \href{https://doi.org/10.3367/UFNe.2020.12.038898}{\emph{Physics Uspekhi}
  {\bfseries 64} (2021) 386}.

\bibitem{bib:Kastor1993}
D.~{Kastor} and J.~{Traschen}, \emph{{Cosmological multi-black-hole
  solutions}}, \href{https://doi.org/10.1103/PhysRevD.47.5370}{\emph{Physical
  Review D} {\bfseries 47} (1993) 5370}
  [\href{https://arxiv.org/abs/hep-th/9212035}{{\ttfamily hep-th/9212035}}].

\bibitem{bib:Wiltshire2013}
D.L.~{Wiltshire}, P.R.~{Smale}, T.~{Mattsson} and R.~{Watkins}, \emph{{Hubble
  flow variance and the cosmic rest frame}},
  \href{https://doi.org/10.1103/PhysRevD.88.083529}{\emph{Physical Review D}
  {\bfseries 88} (2013) 083529}
  [\href{https://arxiv.org/abs/1201.5371}{{\ttfamily 1201.5371}}].

\bibitem{bib:Heinesen2021}
A.~{Heinesen}, \emph{{Redshift drift as a model independent probe of dark
  energy}}, \href{https://doi.org/10.1103/PhysRevD.103.L081302}{\emph{Physical
  Review D} {\bfseries 103} (2021) L081302}
  [\href{https://arxiv.org/abs/2102.03774}{{\ttfamily 2102.03774}}].

\bibitem{bib:Ellis1984}
G.F.R.~{Ellis}, \emph{{Relativistic cosmology: its nature, aims and
  problems.}},  in \emph{General Relativity and Gravitation Conference},
  pp.~215--288, Jan., 1984.

\bibitem{bib:Hertzberg2023}
M.P.~{Hertzberg} and A.~{Loeb}, \emph{{Possible relation between the
  cosmological constant and standard model parameters}},
  \href{https://doi.org/10.1103/PhysRevD.107.063527}{\emph{Physical Review D}
  {\bfseries 107} (2023) 063527}
  [\href{https://arxiv.org/abs/2302.09090}{{\ttfamily 2302.09090}}].

\bibitem{bib:Wistisen2021}
T.N.~Wistisen, C.H.~Keitel and A.D.~Piazza, \emph{Transmutation of protons in a
  strong electromagnetic field},
  \href{https://doi.org/10.1088/1367-2630/abf705}{\emph{New Journal of Physics}
  {\bfseries 23} (2021) 065007}.

\bibitem{bib:On2019}
A.Y.L.~{On}, J.Y.H.~{Chan}, K.~{Wu}, C.J.~{Saxton} and L.~{van
  Driel-Gesztelyi}, \emph{{Polarized radiative transfer, rotation measure
  fluctuations, and large-scale magnetic fields}},
  \href{https://doi.org/10.1093/mnras/stz2683}{\emph{Monthly Notices of the
  Royal Astronomical Society} {\bfseries 490} (2019) 1697}
  [\href{https://arxiv.org/abs/1909.06703}{{\ttfamily 1909.06703}}].

\bibitem{bib:Akahori2016}
T.~{Akahori}, D.~{Ryu} and B.M.~{Gaensler}, \emph{{Fast Radio Bursts as Probes
  of Magnetic Fields in the Intergalactic Medium}},
  \href{https://doi.org/10.3847/0004-637X/824/2/105}{\emph{The Astrophysical
  Journal} {\bfseries 824} (2016) 105}
  [\href{https://arxiv.org/abs/1602.03235}{{\ttfamily 1602.03235}}].

\bibitem{bib:Beck2011}
R.~{Beck}, \emph{{Cosmic Magnetic Fields: Observations and Prospects}},  in
  \emph{25th Texas Symposium on Relativistic AstroPhysics (Texas 2010)},
  F.A.~{Aharonian}, W.~{Hofmann} and F.M.~{Rieger}, eds., vol.~1381 of
  \emph{American Institute of Physics Conference Series}, pp.~117--136, Sept.,
  2011, \href{https://doi.org/10.1063/1.3635828}{DOI}
  [\href{https://arxiv.org/abs/1104.3749}{{\ttfamily 1104.3749}}].

\bibitem{bib:SEF}
P.~Schneider, J.~Ehlers and E.E.~Falco, \emph{{Gravitational Lenses}},
  {Astronomy and Astrophysics Library}, Springer, New York (1992).

\bibitem{bib:Wagner2019}
J.~{Wagner}, \emph{{A Model-Independent Characterisation of Strong
  Gravitational Lensing by Observables}},
  \href{https://doi.org/10.3390/universe5070177}{\emph{Universe} {\bfseries 5}
  (2019) 177} [\href{https://arxiv.org/abs/1906.05285}{{\ttfamily
  1906.05285}}].

\bibitem{bib:Eiroa2002}
E.F.~{Eiroa}, G.E.~{Romero} and D.F.~{Torres}, \emph{{Reissner-Nordstr{\"o}m
  black hole lensing}},
  \href{https://doi.org/10.1103/PhysRevD.66.024010}{\emph{Physical Review D}
  {\bfseries 66} (2002) 024010}
  [\href{https://arxiv.org/abs/gr-qc/0203049}{{\ttfamily gr-qc/0203049}}].

\bibitem{bib:Bozza2002}
V.~{Bozza}, \emph{{Gravitational lensing in the strong field limit}},
  \href{https://doi.org/10.1103/PhysRevD.66.103001}{\emph{Physical Review D}
  {\bfseries 66} (2002) 103001}
  [\href{https://arxiv.org/abs/gr-qc/0208075}{{\ttfamily gr-qc/0208075}}].

\bibitem{bib:Wagner4}
J.~{Wagner}, \emph{{Generalised model-independent characterisation of strong
  gravitational lenses. IV. Formalism-intrinsic degeneracies}},
  \href{https://doi.org/10.1051/0004-6361/201834218}{\emph{Astronomy \&
  Astrophysics} {\bfseries 620} (2018) A86}
  [\href{https://arxiv.org/abs/1809.03505}{{\ttfamily 1809.03505}}].

\bibitem{bib:Wagner6}
J.~{Wagner}, \emph{{Generalised model-independent characterisation of strong
  gravitational lenses - VI. The origin of the formalism intrinsic degeneracies
  and their influence on H$_{0}$}},
  \href{https://doi.org/10.1093/mnras/stz1587}{\emph{Monthly Notices of the
  Royal Astronomical Society} {\bfseries 487} (2019) 4492}
  [\href{https://arxiv.org/abs/1904.07239}{{\ttfamily 1904.07239}}].

\bibitem{bib:Griffiths2021}
R.E.~{Griffiths}, M.~{Rudisel}, J.~{Wagner}, T.~{Hamilton}, P.-C.~{Huang} and
  C.~{Villforth}, \emph{{Hamilton's Object - a clumpy galaxy straddling the
  gravitational caustic of a galaxy cluster: constraints on dark matter
  clumping}}, \href{https://doi.org/10.1093/mnras/stab1375}{\emph{Monthly
  Notices of the Royal Astronomical Society} {\bfseries 506} (2021) 1595}
  [\href{https://arxiv.org/abs/2105.04562}{{\ttfamily 2105.04562}}].

\bibitem{bib:Guetta2005}
D.~{Guetta}, T.~{Piran} and E.~{Waxman}, \emph{{The Luminosity and Angular
  Distributions of Long-Duration Gamma-Ray Bursts}},
  \href{https://doi.org/10.1086/423125}{\emph{The Astrophysical Journal}
  {\bfseries 619} (2005) 412}
  [\href{https://arxiv.org/abs/astro-ph/0311488}{{\ttfamily
  astro-ph/0311488}}].

\bibitem{bib:Vallee2004}
J.P.~{Vall{\'e}e}, \emph{{Cosmic magnetic fields - as observed in the Universe,
  in galactic dynamos, and in the Milky Way}},
  \href{https://doi.org/10.1016/j.newar.2004.03.017}{\emph{New Astronomy
  Reviews} {\bfseries 48} (2004) 763}.

\bibitem{bib:Hilmarsson2021}
G.H.~{Hilmarsson}, D.~{Michilli}, L.G.~{Spitler}, R.S.~{Wharton},
  P.~{Demorest}, G.~{Desvignes} et~al., \emph{{Rotation Measure Evolution of
  the Repeating Fast Radio Burst Source FRB 121102}},
  \href{https://doi.org/10.3847/2041-8213/abdec0}{\emph{The Astrophysical
  Journal Letters} {\bfseries 908} (2021) L10}
  [\href{https://arxiv.org/abs/2009.12135}{{\ttfamily 2009.12135}}].

\bibitem{bib:Zitrin}
A.~{Zitrin}, T.~{Broadhurst}, Y.~{Rephaeli} and S.~{Sadeh}, \emph{{The Largest
  Gravitational Lens: MACS J0717.5+3745 (z = 0.546)}},
  \href{https://doi.org/10.1088/0004-637X/707/1/L102}{\emph{The Astrophysical
  Journal Letters} {\bfseries 707} (2009) L102}
  [\href{https://arxiv.org/abs/0907.4232}{{\ttfamily 0907.4232}}].

\bibitem{bib:Aluri2022}
P.~{Kumar Aluri}, P.~{Cea}, P.~{Chingangbam}, M.-C.~{Chu}, R.G.~{Clowes},
  D.~{Hutsem{\'e}kers} et~al., \emph{{Is the observable Universe consistent
  with the cosmological principle?}},
  \href{https://doi.org/10.1088/1361-6382/acbefc}{\emph{Classical and Quantum
  Gravity} {\bfseries 40} (2023) 094001}
  [\href{https://arxiv.org/abs/2207.05765}{{\ttfamily 2207.05765}}].

\bibitem{bib:Hackmann2008}
E.~{Hackmann}, V.~{Kagramanova}, J.~{Kunz} and C.~{L{\"a}mmerzahl},
  \emph{{Analytic solutions of the geodesic equation in higher dimensional
  static spherically symmetric spacetimes}},
  \href{https://doi.org/10.1103/PhysRevD.78.124018}{\emph{Physical Review~D}
  {\bfseries 78} (2008) 124018}
  [\href{https://arxiv.org/abs/0812.2428}{{\ttfamily 0812.2428}}].

\end{thebibliography}\endgroup

% Please avoid comments such as "For a review'', "For some examples",
% "and references therein" or move them in the text. In general,
% please leave only references in the bibliography and move all
% accessory text in footnotes.

% Also, please have only one work for each \bibitem.

\end{document}